\documentclass[aps,prc,twocolumn,floatfix,superscriptaddress]{revtex4}
\usepackage{epsfig}
\usepackage{amsmath}
\usepackage{color}

\usepackage{graphicx}

\begin{document}

\title{Anisotropic flow of thermal photons at RHIC and LHC}
\author{Rupa Chatterjee}
\email{rupa@vecc.gov.in}
\author{Pingal Dasgupta}
\author{Dinesh K. Srivastava}
\affiliation{Variable Energy Cyclotron Centre, HBNI, 1/AF, Bidhan Nagar, Kolkata-700064, India}

\begin{abstract}
We calculate elliptic and triangular flow parameters of thermal photons using an event-by-event hydrodynamic model with fluctuating initial conditions at 200A GeV Au+Au collisions at RHIC and at 2.76A TeV Pb+Pb collisions at the LHC for three different centrality bins.   
The photon elliptic flow shows strong centrality dependence where $v_2(p_T)$ increases towards peripheral collisions both at RHIC and at the LHC energies. However, the triangular flow parameter does not show significant dependence to the collision centrality. The elliptic as well as the triangular flow parameters found to underestimate the PHENIX data at RHIC by a large margin for all three centrality bins. We calculate $p_T$ spectrum and anisotropic flow of thermal photons from 200A GeV Cu+Cu collisions at RHIC for 0--20\% centrality bin  and compare with the results with those from Au+Au collisions. The production of thermal photon is found to decrease significantly for Cu+Cu collisions compared to  Au+Au collisions. However, the effect of initial state fluctuation is found to be more pronounced for anisotropic flow resulting in larger $v_2$ and $v_3$ for Cu+Cu collisions.
We study the correlation between the anisotropic flow parameters and the corresponding initial spatial anisotropies from their event by event distributions at RHIC and at the LHC energies. The linear correlation between  $v_2$ and $\epsilon_2$ is found be stronger compared to the correlation between $v_3$ and $\epsilon_3$. In addition, the correlation coefficient is found to be larger at LHC than at  RHIC. 

\end{abstract}

\maketitle

\section{Introduction} 
Hydrodynamic model with event-by-event (e-by-e) fluctuating initial conditions is considered to be one of the most successful frameworks in recent times to study the evolution of the hot and dense Quark-Gluon-Plasma (QGP) produced in collisions of heavy nuclei at the relativistic energies~\cite{hannu, hama, andrade, bjorn, pt, scott, hannah, v3_uli, flow_lhc}.
Hydrodynamic model with smooth initial density distribution was used for a long time  earlier to explain the particle spectra and large elliptic flow of hadrons produced in  heavy ion collisions at the Relativistic Heavy Ion Collider (RHIC)~\cite{uli}. However, recent studies have shown that hydrodynamics using fluctuating initial conditions explain the the elliptic flow of charged particles even for  most central collisions and also the significantly large  triangular flow  of hadron both at RHIC and at the LHC energies. These were underestimated earlier by hydrodynamic model calculation using smooth initial density distribution.

Electromagnetic radiations are considered as one of the promising probes to study the properties of QGP formed in relativistic heavy ion collisions~\cite{phot}.  Thermal photons calculated using e-by-e hydrodynamics along with prompt photons from next-to-leading (NLO) order pQCD calculation found to explain the  direct photon data both at RHIC and at the LHC in the region $p_T >$ 2 GeV/$c$~\cite{chre3, crs, alice_phot}. However, it is to be noted that for $p_T <$ 2 GeV/$c$, the region  dominated by radiation from the hadronic matter phase is still remain unexplained by theory calculation. 

 It has been shown earlier that fluctuations in the initial density distribution enhance the production of thermal photons compared to a smooth initial state in the hydrodynamic model calculation significantly~\cite{chre1}. This enhanced production results in a better description of experimental direct photon spectra in the $p_T$ region 2--4 GeV/$c$. In addition, the enhancement due to fluctuations in the initial state  is found to be more for peripheral collisions than for central collisions~\cite{chre2} and  for a fixed centrality bin, the photon production is found to be more at RHIC than at the LHC energy. 

Fluctuations in the initial density distribution increases the elliptic flow (in the region $p_T$ 2 to 5 GeV/$c$)~\cite{chre3} and also result in a significantly large triangular flow of thermal photons~\cite{crs}. 

However, the elliptic flow calculated using e-by-e hydrodynamic model does not improve the poor agreement of the direct photon $v_2$ data at RHIC. The data remain unexplained even with many recent studies with sophisticated model calculation~\cite{v2_1,v2_2, v2_3, v2_4}. 
 This is known as the photon $v_2$ puzzle.

The PHENIX collaboration  has recently reported the elliptic and triangular flow of  direct photons from 200A GeV Au+Au collisions at RHIC at centrality bins 0--20\%, 20--40\% and 40--60\%~\cite{phenix_phot}. The direct photon spectra from 2.76A TeV Pb+Pb collisions by ALICE Collaboration are also available now for centrality bins 0--20\%, 20--40\% and 40--80\%~\cite{alice_phot}.
Thus, calculation of photon anisotropic flow parameters at different collision centralities at RHIC and at the LHC energies and comparison with experimental data would be valuable and might help us to understand the photon $v_2$ puzzle.

We calculate the elliptic and triangular flow parameters of thermal photons for three different centrality bins using e-by-e hydrodynamic model from 200A GeV Au+Au collisions at RHIC and compare the results with PHENIX experimental data. We see that both  $v_2$ and $v_3$ at RHIC underestimate the data by a large margin for all three centralities.  The photon anisotropic flow parameters at LHC energy are also calculated for different centrality bins.  The elliptic flow parameter from hydrodynamic model calculation is found to increase towards peripheral collision, however, the triangular flow parameter does not show strong centrality dependence. 

In addition, we calculate the $p_T$ spectrum, elliptic and triangular flow of thermal photon for a smaller system produced in Cu+Cu collisions at 200A GeV at RHIC for 0--20\% centrality bin and compare the results  with  those from Au+Au collisions. The presence of fluctuations in the initial density distribution results in a larger elliptic as well as triangular flow of photon for Cu+Cu collisions than for Au+Au collisions. 
Next we study the correlation between the photon anisotropic flow parameters ($v_n$) and their corresponding initial eccentricities ($\epsilon_n$) from their e-by-e distribution at different collision centralities both at RHIC and at the LHC. The  correlation co-efficient, which is a measures of the strength of the linear association between two variables ($v_n$ and $\epsilon_n$) is found to be larger for elliptic flow than for the triangular flow parameter. It is also found to larger at LHC energy than at RHIC.

\section{Initial conditions and hydrodynamic framework }
We have used a (2+1) dimensional ideal hydrodynamic model framework with fluctuating initial conditions  developed in~\cite{hannu}  to calculate the spectra, elliptic, and triangular flow anisotropies of thermal photons  at RHIC and at the LHC energies. This model has been used successfully earlier to explain the $p_T$ spectra and elliptic flow of hadrons at RHIC~\cite{hannu} for most central collisions and also  been used to calculate the photon production at the RHIC and at the LHC energies~\cite{chre1,chre2,chre3}. 

In the Monte Carlo Glauber initial state a standard two-parameter Woods-Saxon  nuclear density profile is used to randomly distribute the nucleons into the two colliding nuclei. 
We take  the inelastic nucleon nucleon cross-section $\sigma_{NN}=$ 42 mb and 64 mb at RHIC and LHC respectively. 

A 2-dimensional Gaussian distribution function of the form 
\begin{equation}
  s(x,y) = \frac{K}{2 \pi \sigma^2} \sum_{i=1}^{\ N_{\rm WN}} \exp \Big( -\frac{(x-x_i)^2+(y-y_i)^2}{2 \sigma^2} \Big).
 \label{eq:eps}
\end{equation}
is used to distribute initial entropy density around the participants in a wounded nucleon profile. K is a constant in the Eq. above fixed from final charged particle multiplicity  and the position of the $i$th nucleon in the transverse plane is denoted by ($x_i, y_i$). 
The granularity or the size of the initial density fluctuation is decide by the parameter $\sigma$.
It is a free parameter and we use a default value of  $\sigma=$ 0.4 fm for our calculation as before~\cite{hannu,chre1,chre2,chre3}. The initial formation time $\tau_0=$ 0.17 and 0.14 fm/$c$ for Au+Au collisions at RHIC and for Pb+Pb collisions at the  LHC respectively are taken  from EKRT minijet saturation model~\cite{ekrt}. The initial formation time for Cu+Cu collisions at 200A GeV  is taken as 0.2 fm/$c$. 

We use complete leading order (LO) plasma rates from~\cite{amy} to calculate the photon production from the QGP phase. It has been shown earlier that the  (NLO) plasma rates~\cite{nlo_thermal} do not contribution significantly to the photon production for $p_T >$ 2 GeV/$c$. The hadronic phase contribution is estimated using the parameterization given in Ref.~\cite{trg}. 

The temperature at freeze-out is taken as 160 MeV which reproduces the measured $p_T$ spectra of charges pions at RHIC and at the LHC. 170 MeV is considered as the transition temperature  from the plasma phase to hadronic phase and  a lattice based equation of state~\cite{eos} for our calculation. 

 The total thermal emission  is calculated  by integrating the emission rates ($R=EdN/d^3pd^4x$) over the space-time history using the relation:
\begin{equation}
E \frac{dN}{d^3p}= \int d^4x \, R \left(E^*(x),T(x)\right).
\end{equation}
T(x) is the local temperature in the equation above.  $E^* (x)$ = $p^\mu u_\mu (x)$ where,  $p^\mu$ is the four-momentum of the photons and $u_\mu$ is the local four-velocity of the flow field.
The anisotropic flow co-efficients $v_n$ are estimated by expanding the invariant particle distribution in transverse plane using Fourier decomposition:

\begin{equation}\label{eq: v2}
 \frac{dN}{d^2p_TdY} = \frac{1}{2\pi} \frac{dN}{ p_T dp_T dY}[1+ 2\, \sum_{n=1}^{\infty} v_n (p_T) \, \rm{cos} \, n (\phi - \psi_n^{PP})] \, .
\end{equation}
%
The anisotropic flow parameters are calculated with respect to the the participant plane angle $\psi_n^{\rm PP}$ where,
\begin{equation}
  \psi_{n}^{\text{PP}} = \frac{1}{n} \arctan 
  \frac{\int \mathrm{d}x \mathrm{d}y \; r^2 \sin \left( n\phi \right) \varepsilon\left( x,y,\tau _{0}\right) } 
  { \int \mathrm{d}x \mathrm{d}y \; r^2 \cos \left( n\phi \right) \varepsilon\left( x,y,\tau _{0}\right)}  + \pi/n \, .
\end{equation}
The initial eccentricities are estimated using the relation:
\begin{equation}
  \epsilon_{n} = -\frac{\int \mathrm{d} x \mathrm{d} y \; r^{2} \cos \left[
  n\left( \phi -\psi_{n}^{\text{PP}}\right) \right] \varepsilon \left(
  x,y,\tau_{0}\right) } {\int \mathrm{d} x \mathrm{d} y \; r^{2} \varepsilon
  \left( x,y,\tau _{0}\right) } \, . 
\end{equation}

Here $\varepsilon$ is the energy density, $r^{2}=x^{2}+y^{2}$, and $\phi$ is
the azimuthal angle.

\begin{figure}
\centerline{\includegraphics*[width=8.0 cm]{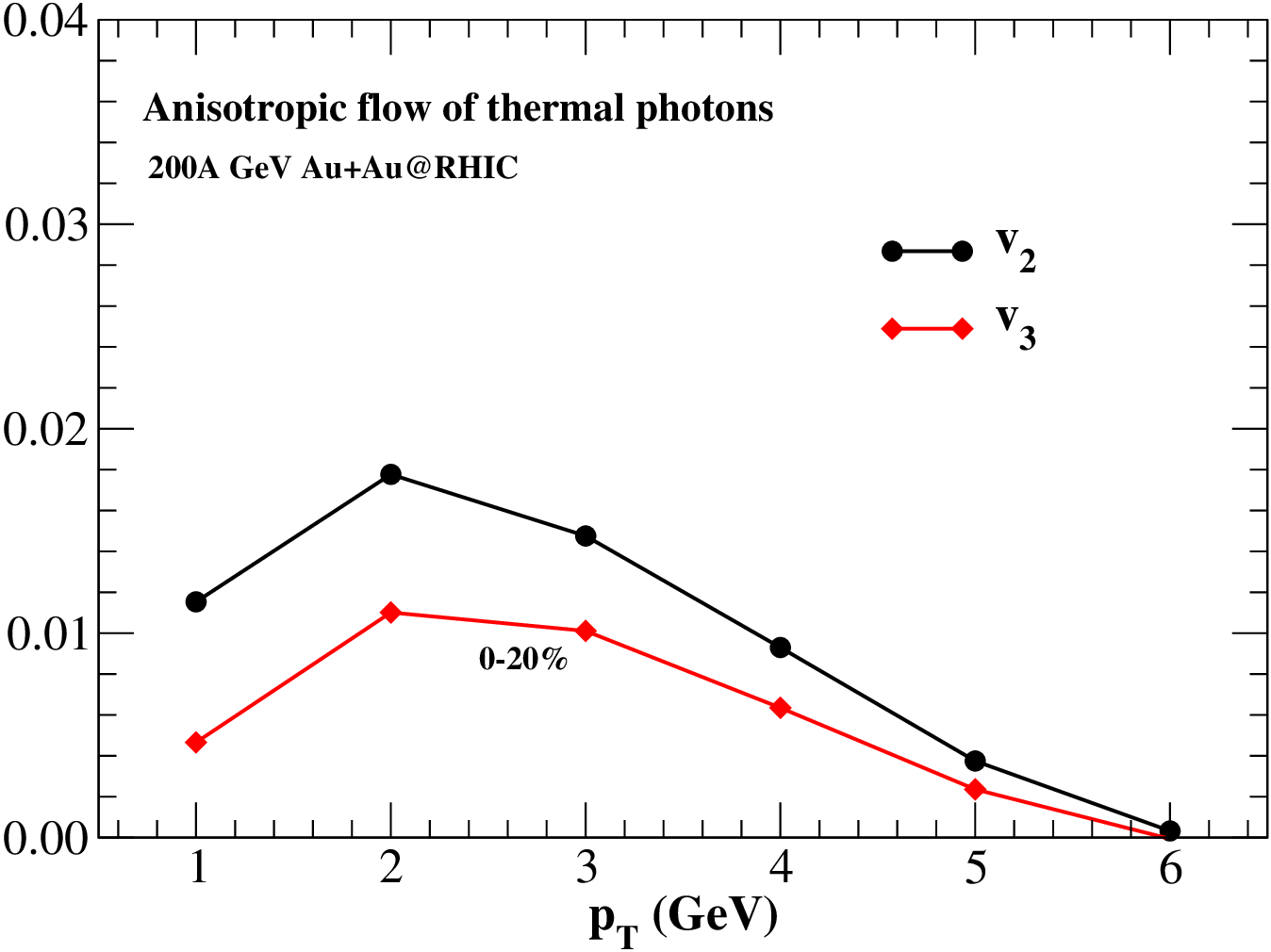}}
\centerline{\includegraphics*[width=8.0 cm]{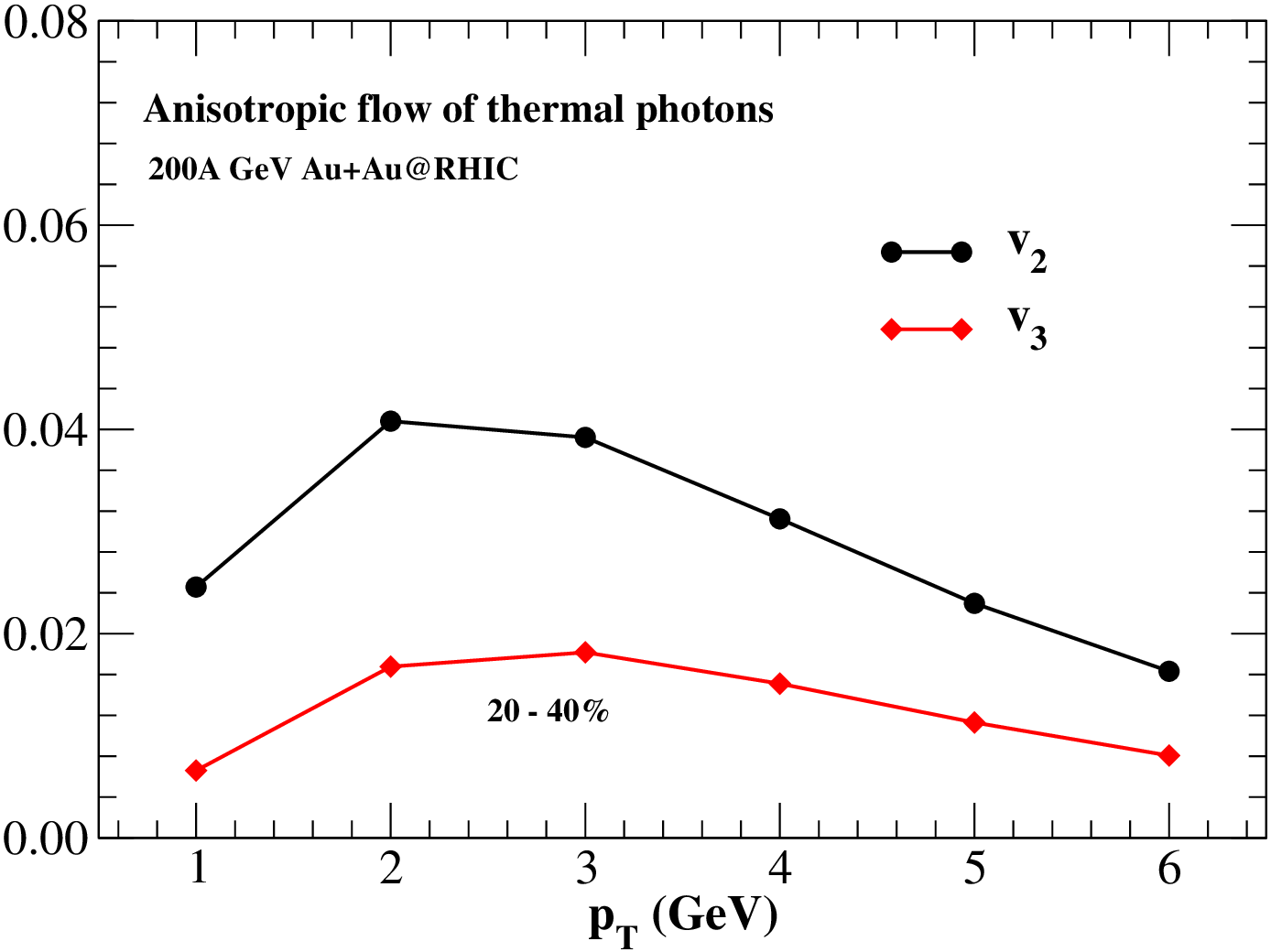}}
\centerline{\includegraphics*[width=8.0 cm]{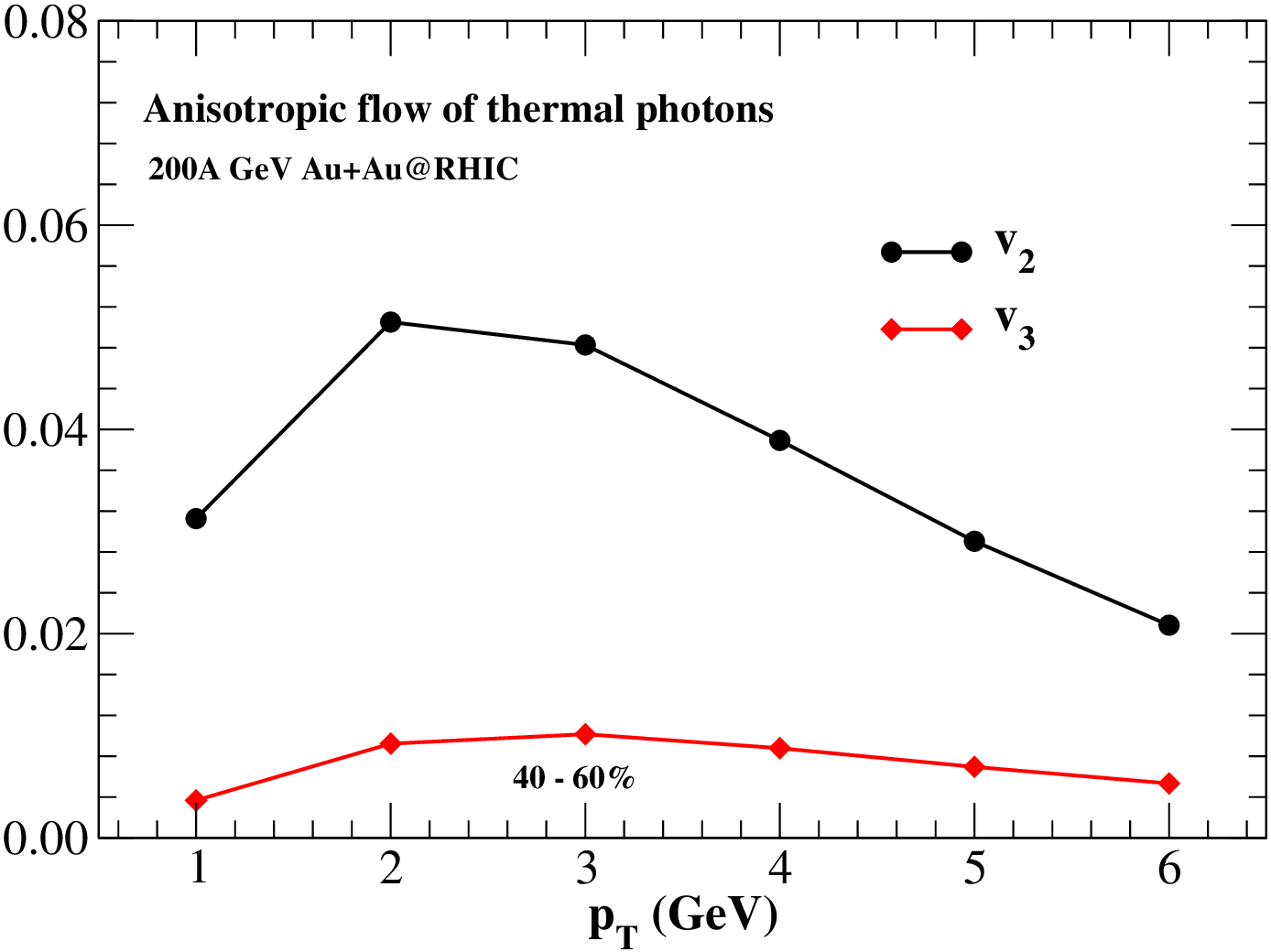}}
\caption{(Color online) Elliptic  and triangular flow parameter of thermal photons for 200A GeV Au+Au collisions at RHIC and for centrality bins 0--20\%, 20--40\%, and 40--60\%. }
\label{fig1}
\end{figure}

\begin{figure}
\centerline{\includegraphics*[width=8.0 cm]{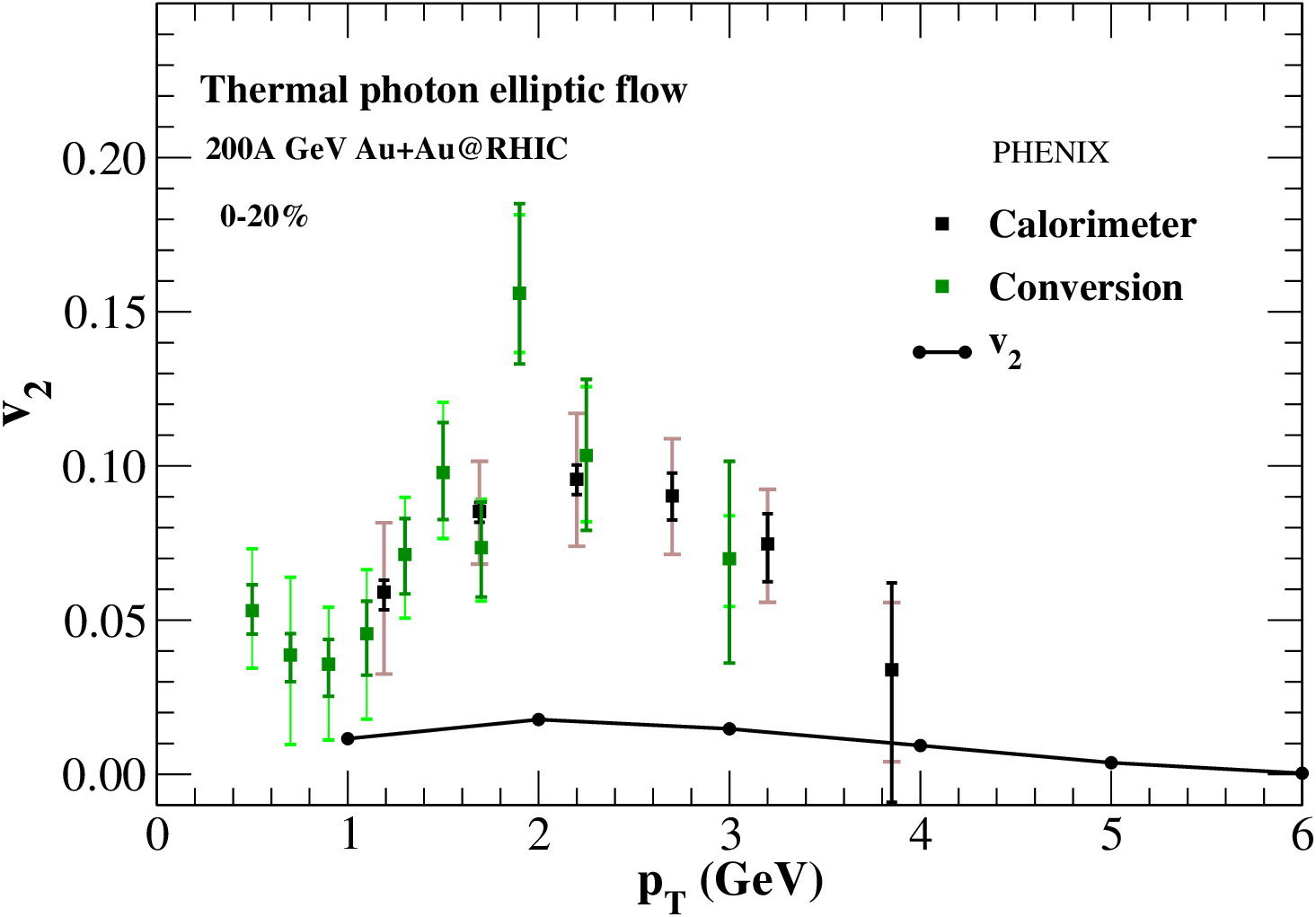}}
\centerline{\includegraphics*[width=8.0 cm]{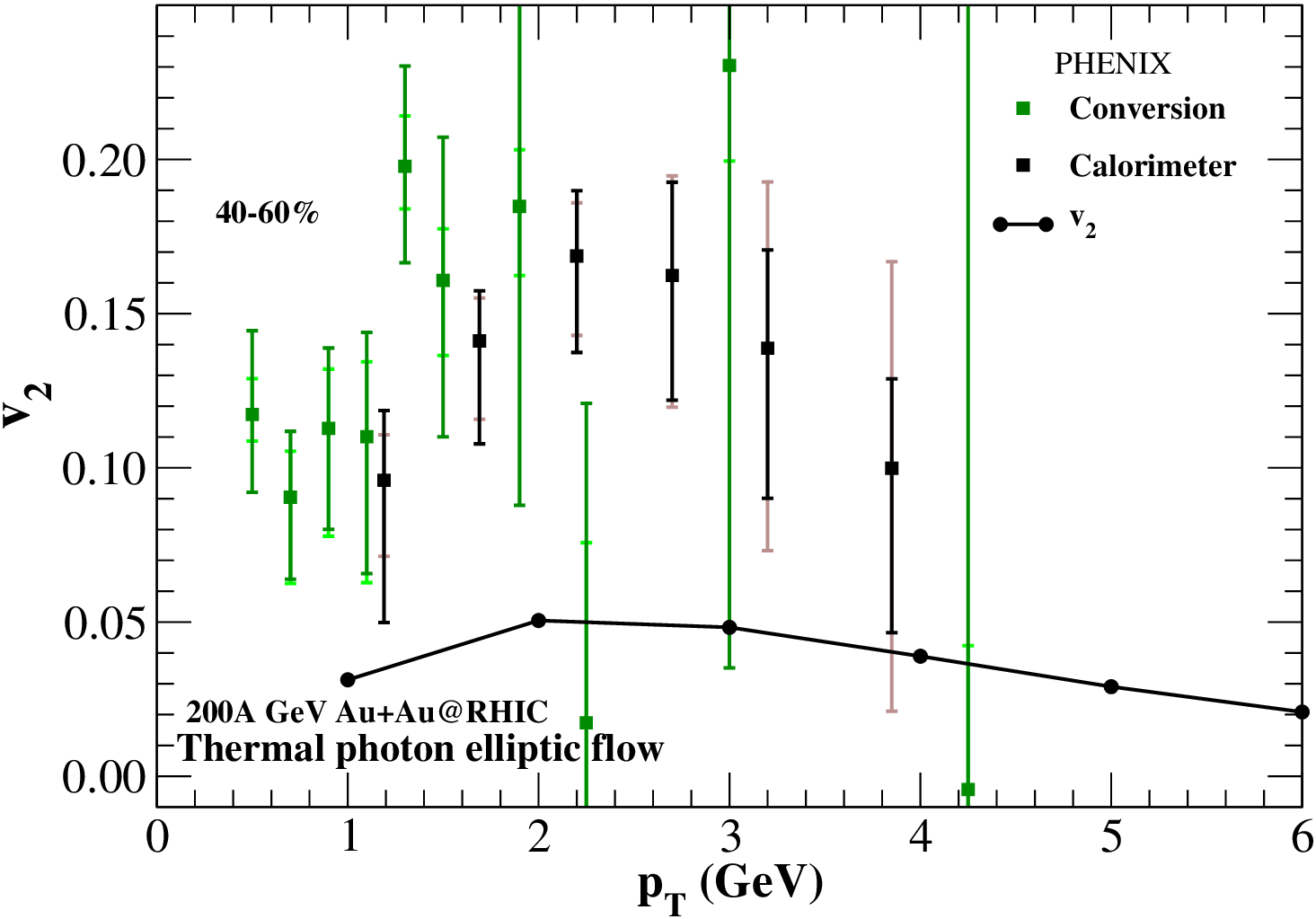}}
\caption{(Color online) Elliptic flow of thermal photons at RHIC and PHENIX direct photon $v_2$ data~\cite{phenix_phot}.}
\label{fig2}
\end{figure}

\section{Results}
\subsection{Au+Au collisions at 200A GeV at RHIC}
The elliptic and triangular flow parameters of thermal photons for three different centrality bins at RHIC are shown in Fig.~\ref{fig1}. The anisotropic flow parameters for each centrality bin is obtained by taking average of results over 400 random events with fluctuating initial conditions. As hadrons are emitted from the surface of freeze-out, one needs a much larger number of events to calculate the hadronic observables. However,  photons are emitted from the entire space time 4-volume of the produced system and those with $p_T >$ 1 GeV/$c$ are mostly from the interior of the expanding fireball. Thus, averaging over 400 events are found to be sufficient for photon anisotropic flow calculation using e-by-e fluctuating initial conditions. 

We see that the $v_2$ and $v_3$ of thermal photons as a function of $p_T$ are small and close to each other for 0--20\% centrality bin. The $v_2(p_T)$ for 20--40\% centrality bin is significantly larger than the same at 0--20\% centrality bin. However, the $v_3(p_T)$ increases only slightly for 20--40\% central collisions. We see $v_2(p_T)$ is largest for 40--60\% centrality bin. On the other hand, $v_3(p_T)$ is found to be slightly smaller for 40--60\% centrality bin than the triangular flow parameter calculated  for 20--40\% centrality bin. 

One can see that the elliptic flow of thermal photons calculated using e-by-e hydrodynamic model shows clear dependence on the collision centrality and the value of $v_2(p_T)$ is larger for peripheral collisions than for central collisions. We have seen this centrality dependence in earlier study as well, using smooth initial conditions (see Fig 4 of ~\cite{cfhs}). The initial spatial anisotropy is larger for peripheral collisions than for central collisions. In addition, the relative contribution of the hadronic phase compared to the QGP phase to the photon $v_2$ increases for peripheral collisions. The fluctuations in the initial density distribution also increases the elliptic flow more for peripheral collision and as a result we see larger $v_2(p_T)$ for peripheral collisions. 

For thermal photons the contribution to $v_3$ comes from both QGP as well as from the hadronic phase. We have also seen that initial geometry is important for non-zero photon $v_3$, at the same time the fluctuation driven larger transverse flow velocity plays the most significant role in determining the triangular flow of photons~\cite{crs}.
However, the $v_3(p_T)$ do not show significant centrality dependence. Earlier studies have shown that hadronic $v_3$ which originates  due to the fluctuations in the initial density distribution,  also does not depend on the collision centrality.

\begin{figure}
\centerline{\includegraphics*[width=8.0 cm]{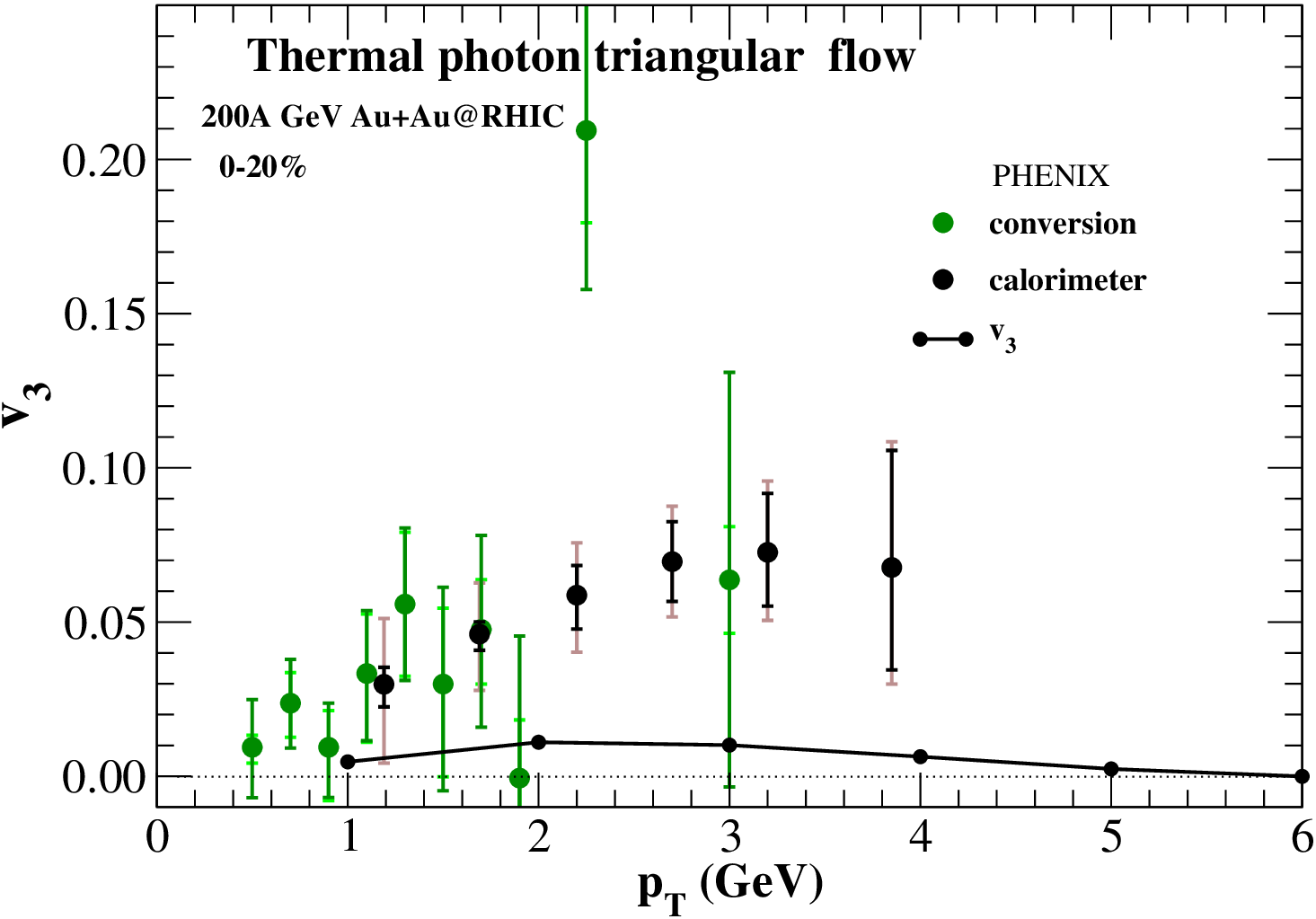}}
\centerline{\includegraphics*[width=8.0 cm]{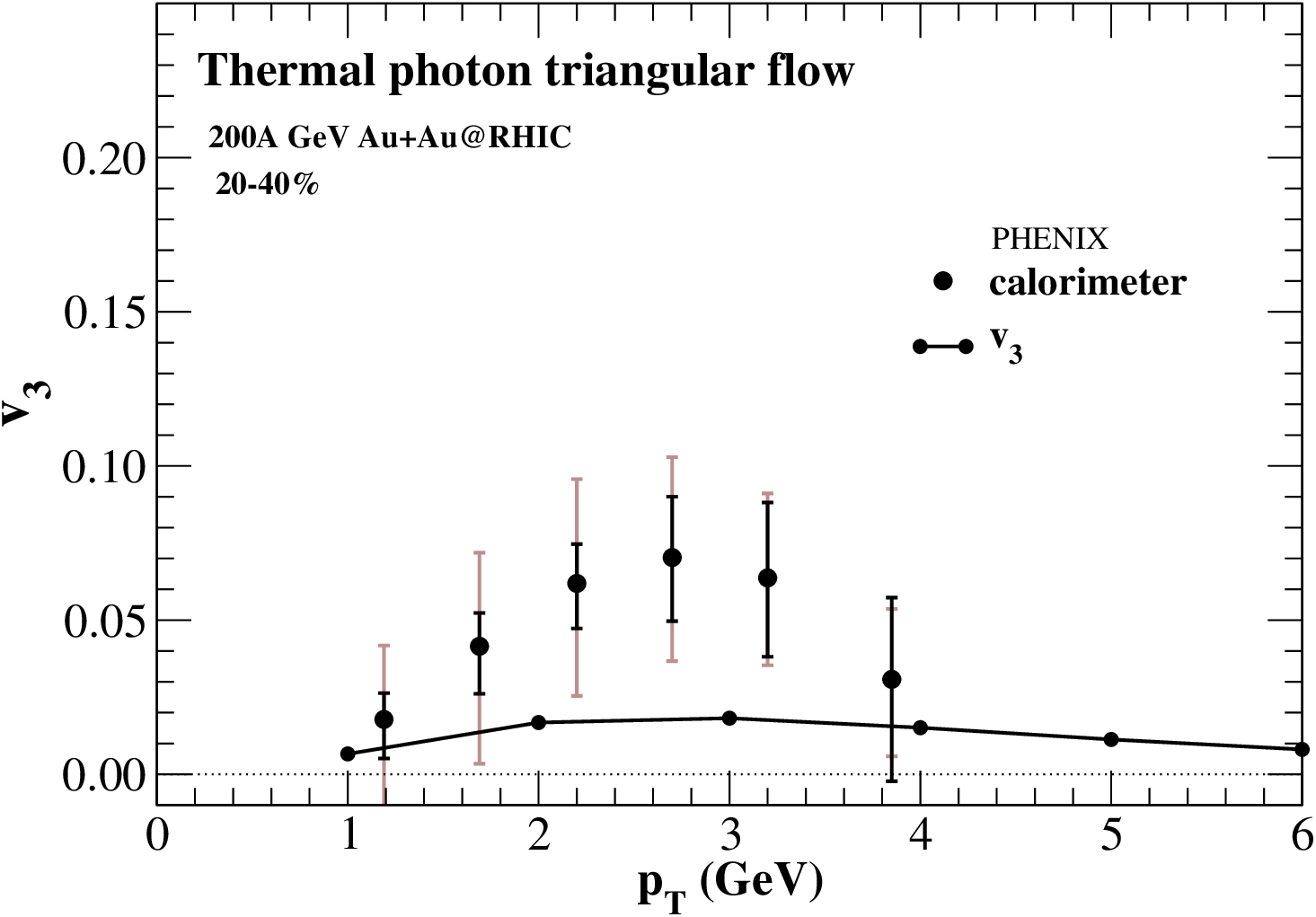}}
\centerline{\includegraphics*[width=8.0 cm]{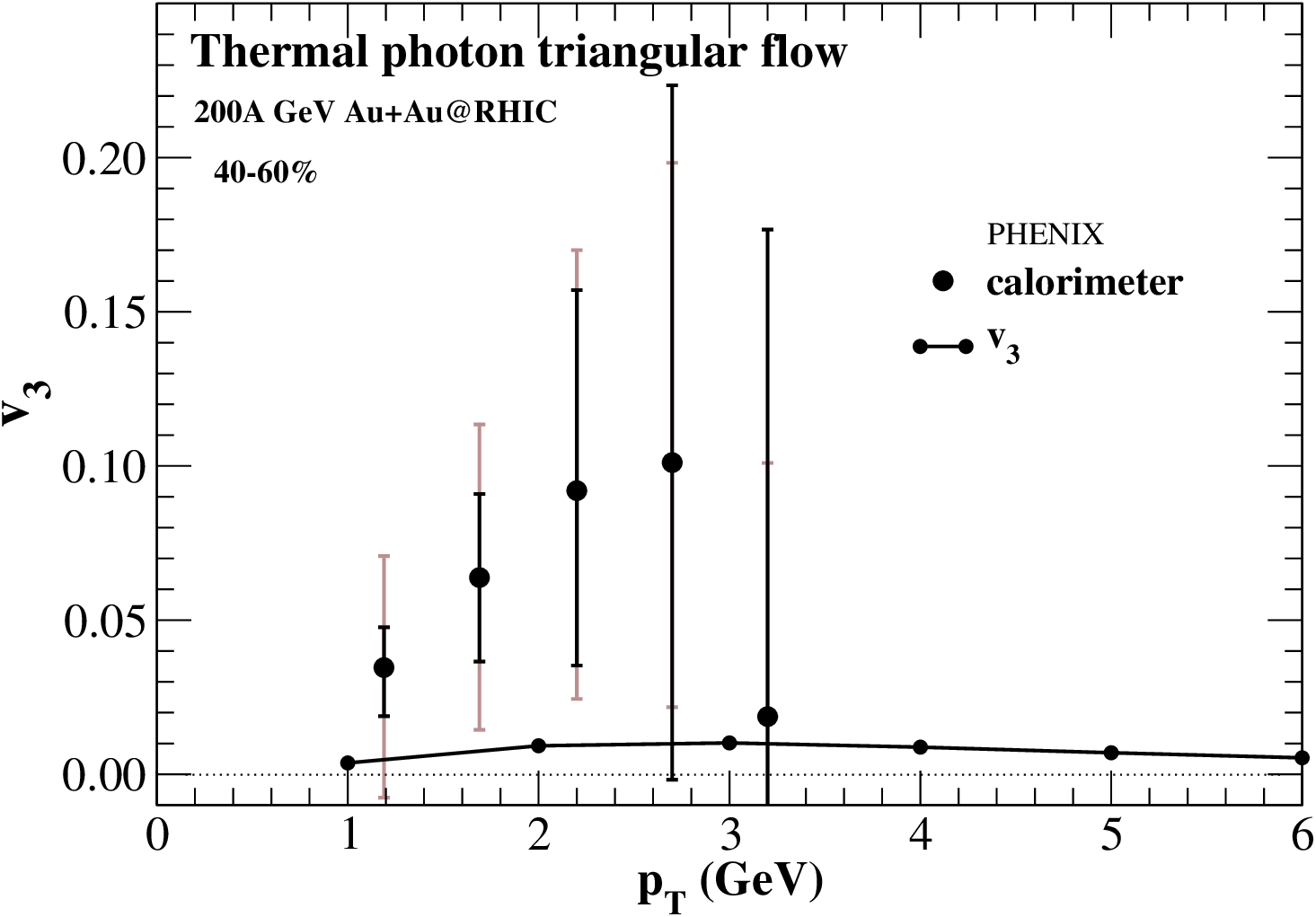}}
\caption{(Color online) Triangular flow of thermal photons for three different centrality bins at RHIC and PHENIX direct photon $v_3$ data~\cite{phenix_phot}.}
\label{fig3}
\end{figure}

The comparison of the elliptic and triangular flow parameters calculated using hydrodynamical model with the PHENIX anisotropic flow  data is shown in Fig~\ref{fig2} and Fig.~\ref{fig3} [comparison of photon $v_2$ for 20--40\% centrality with PHENIX data  is shown earlier in Ref.~\cite{chre2}]. Our results under-predict the data by a large margin both for   elliptic and triangular flow parameters and also for all the centrality bins shown here. Thus, we see that a photon $v_3$ puzzle is also there. It is to be noted that by looking at the experimental data it is difficult to draw any conclusions conclude anything about the centrality dependence of direct photon $v_2$ and $v_3$.
\subsection{Pb+Pb collisions at 2.76A TeV at LHC}
\begin{figure}
\centerline{\includegraphics*[width=8.0 cm]{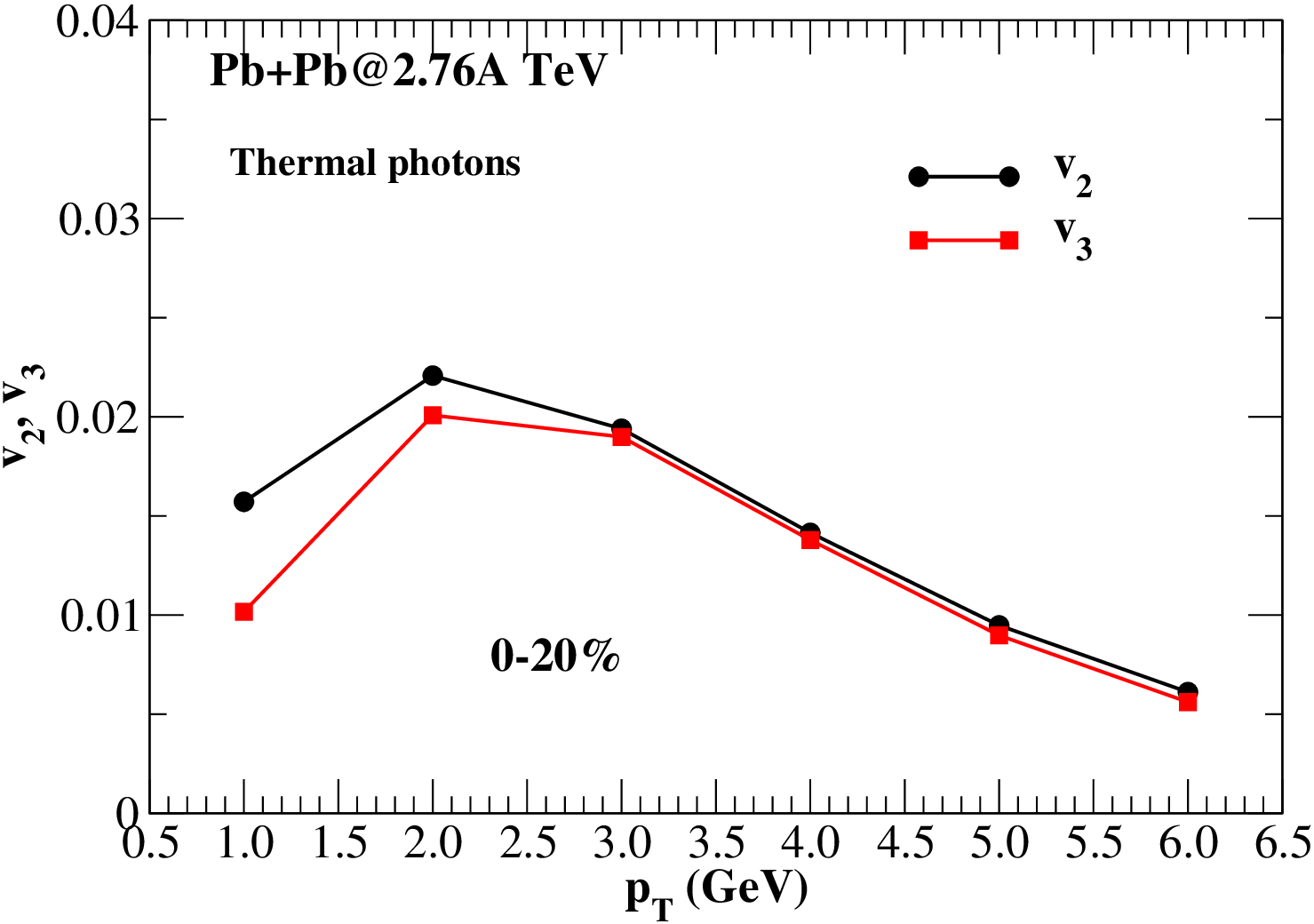}}
\centerline{\includegraphics*[width=8.0 cm]{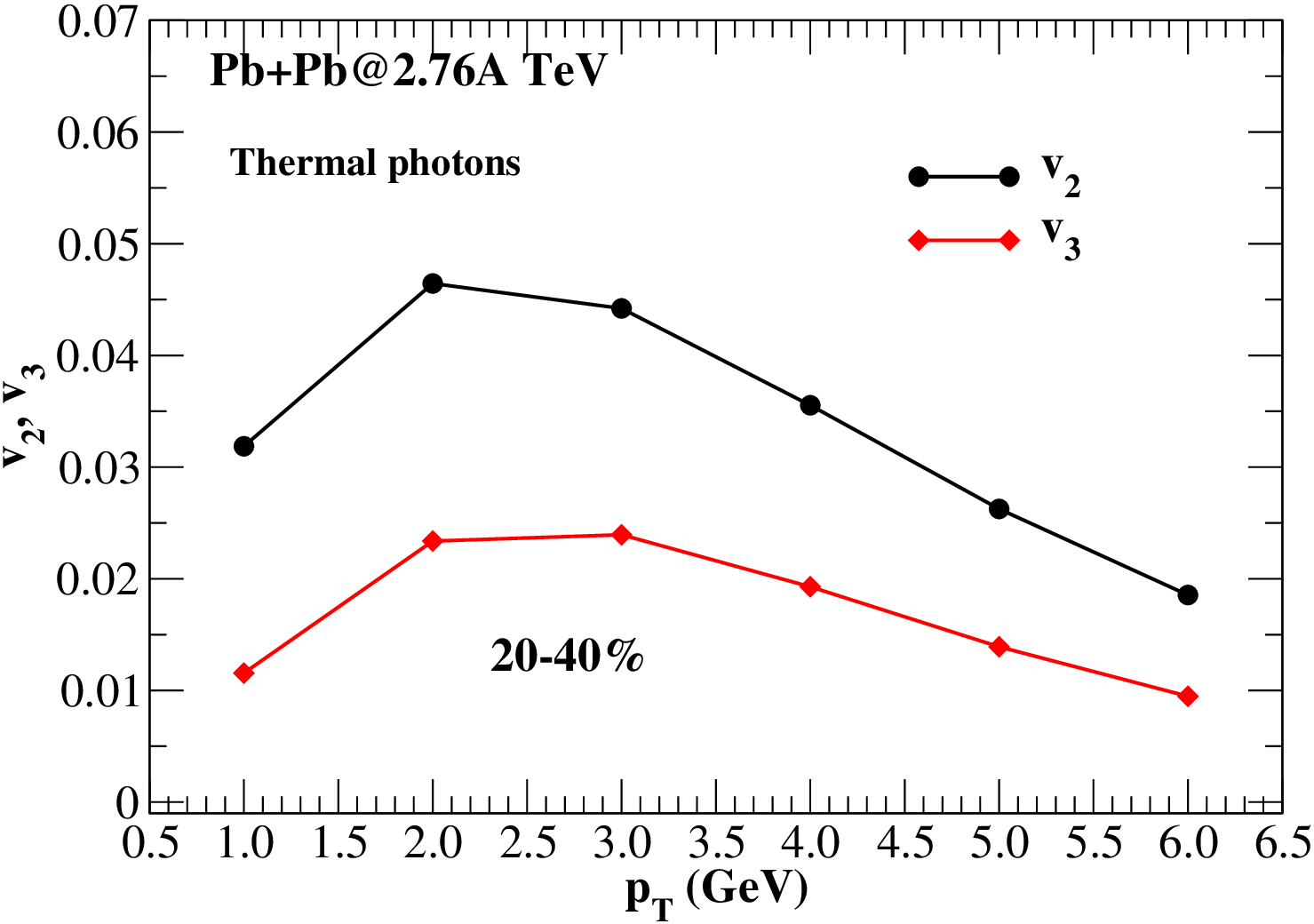}}
\centerline{\includegraphics*[width=8.0 cm]{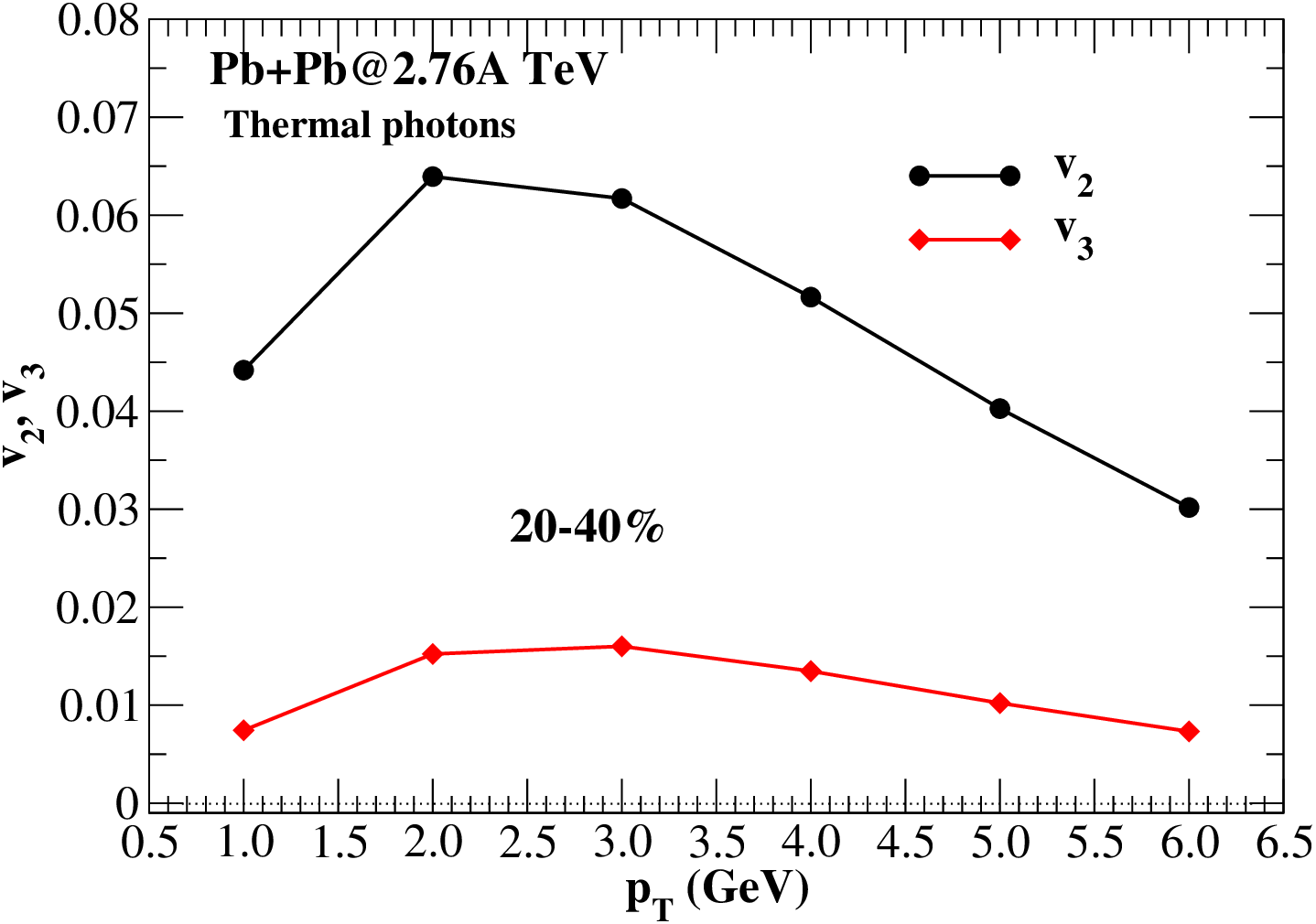}}
\caption{(Color online) Elliptic and triangular flow of thermal photons for three different centrality bins at 2.76A TeV Pb+Pb collisions at LHC. }
\label{fig4}
\end{figure}

Thermal photon $v_2$ and $v_3$ from 2.76A TeV Pb+Pb collisions at LHC and for centrality bins 0--20\%, 20--40\% and 40--60\% are shown in Fig.~\ref{fig4}. Similar to RHIC, we see that the elliptic and triangular flow parameters for 0--20\% central collision are close and  almost on top of each other in the region $p_T > $ 2 GeV/$c$. The $v_2(p_T)$ is shown to rise with collision centrality. In addition, the elliptic flow result for a particular centrality bin is found be slightly larger at LHC than at the RHIC energy. The $v_3$ results are found to be small and do not show strong centrality dependence.

\subsection{Cu+Cu collisions at RHIC}
We have seen earlier that the effect of fluctuations in the initial density distribution is more pronounced for peripheral collisions and for smaller systems~\cite{chre2}. 
The PHENIX Collaboration  has reported the ratio of direct and inclusive photon yield for various systems including Cu+Cu, for minimum bias collisions at RHIC~\cite{cu_phenix}. One can expect to have direct photon data for different centrality bins of Cu+Cu collisions as well in near future. 
Thus, we calculate the spectra and anisotropic flow parameters of thermal photons from Cu+Cu collisions at RHIC. We consider 0--20\% central events only keeping in mind that result from hydrodynamical model is reliable for the most central collisions of Cu nuclei. 

\begin{figure}
\centerline{\includegraphics*[width=8.0 cm]{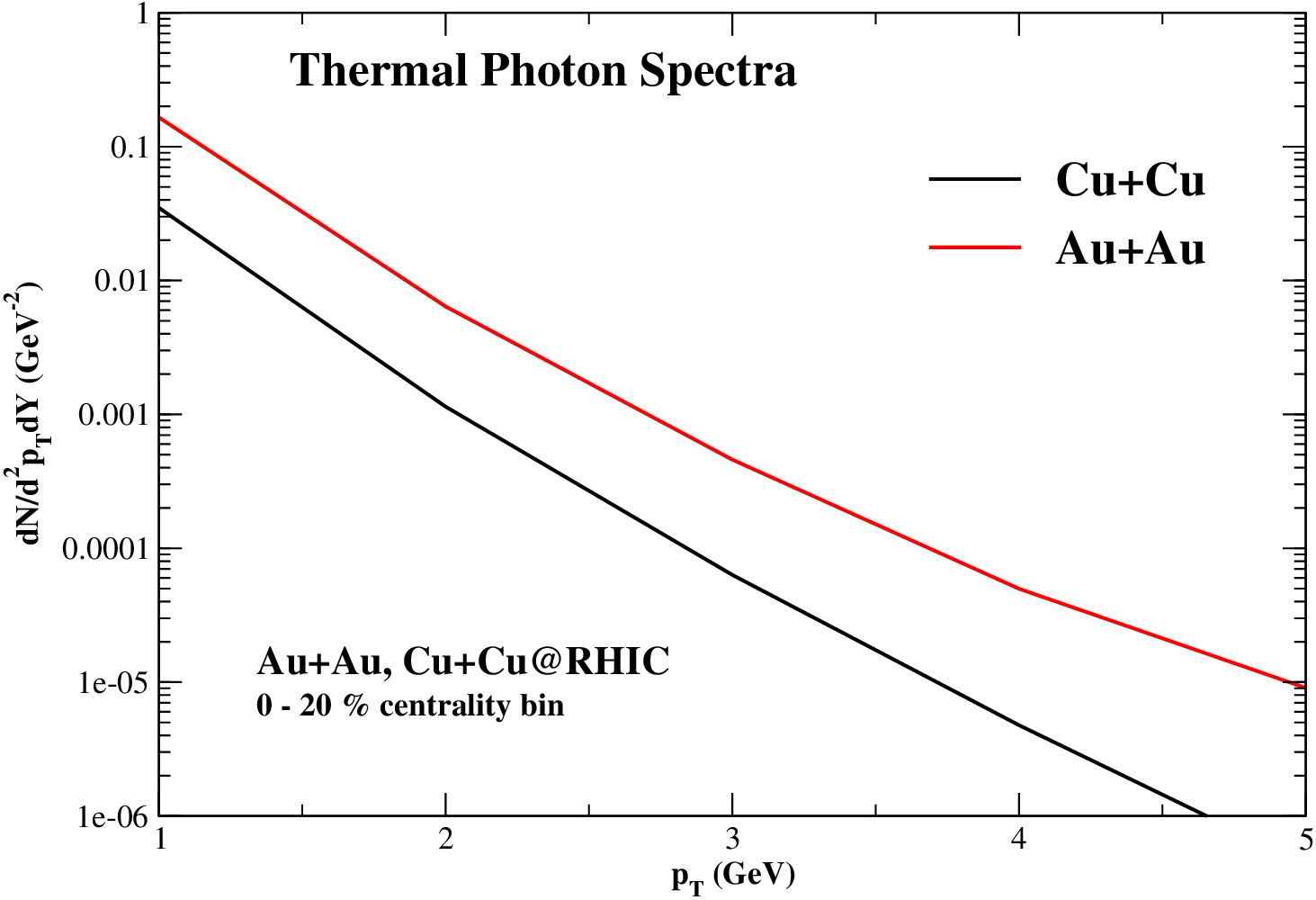}}
\centerline{\includegraphics*[width=8.0 cm]{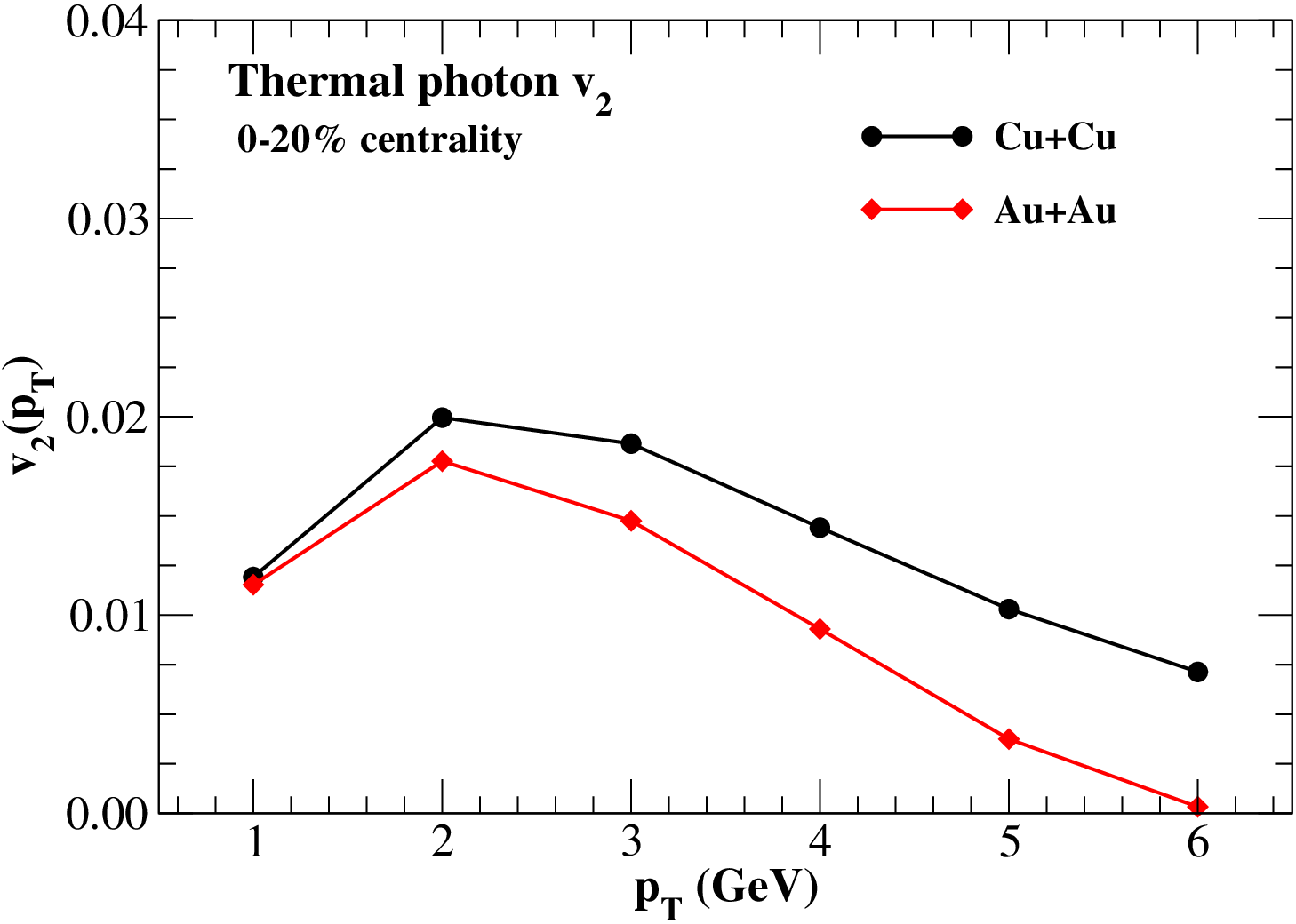}}
\centerline{\includegraphics*[width=8.0 cm]{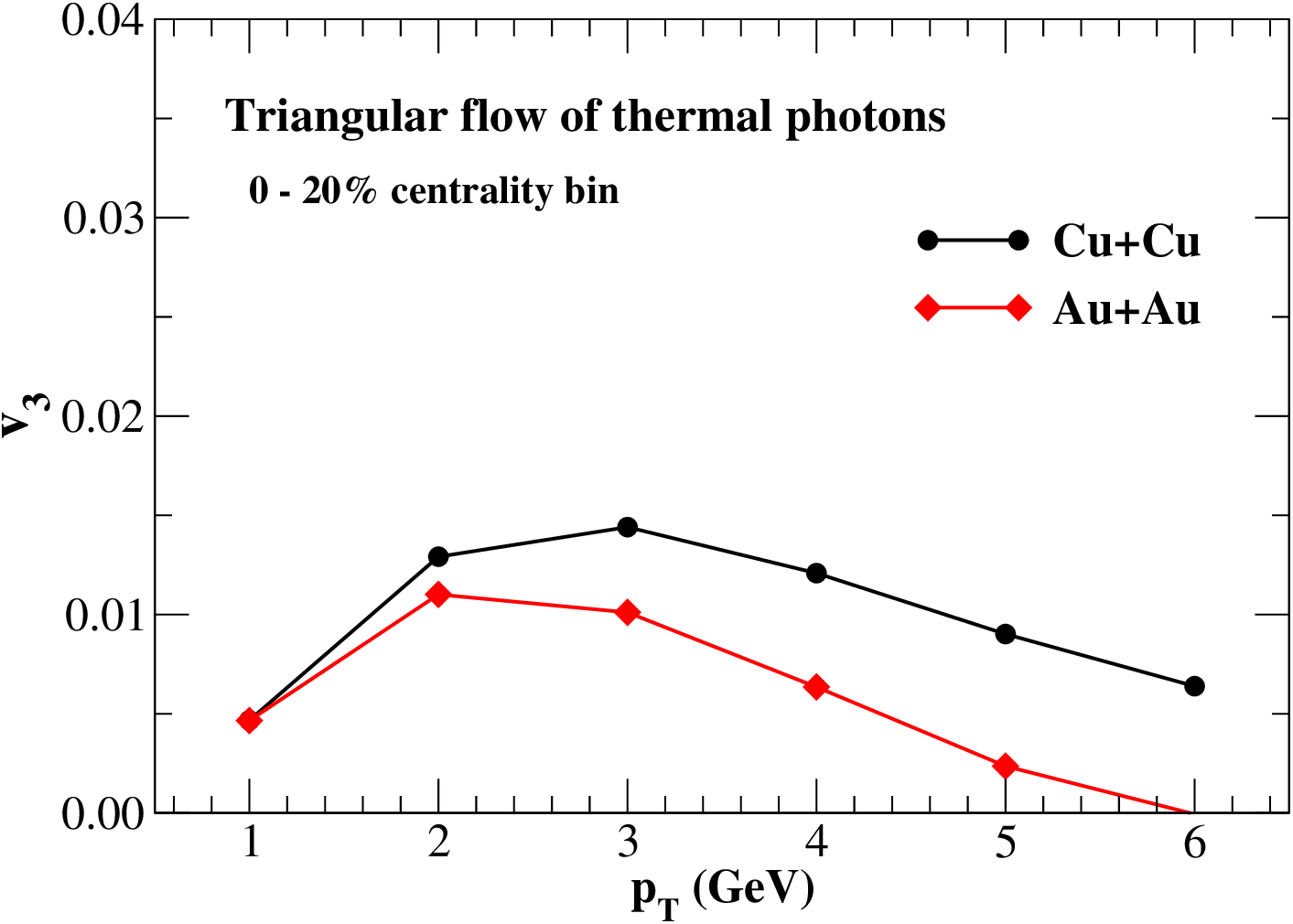}}
\caption{(Color online) The $p_T$ spectra, elliptic and triangular flow parameters of thermal photons from 200A GeV Cu+Cu and Au+Au collisions at RHIC and for 0--20\% centrality bin. }
\label{fig5}
\end{figure}

The thermal photon spectra, elliptic and triangular flow parameters  from 200A GeV Cu+Cu collisions at RHIC and comparison with the corresponding results from Au+Au collisions  are shown in Fig~\ref{fig5}. 
The $p_T$ spectrum is found to be about a factor of 5--10 smaller for Cu+Cu collisions than for the Au+Au collisions as the system produced in collisions of Cu nuclei has much smaller energy density as well as temperature than the one produced in collisions of Au nuclei.
 
One can see that the anisotropic flow calculated for Cu+Cu collisions are larger compared to the results from Au+Au collisions in the region $p_T >$ 2 GeV/$c$. This is due the fluctuations in the initial density distribution that give rise to larger flow parameters for Cu+Cu collisions.

\begin{figure}
\centerline{\includegraphics*[width=7.0 cm]{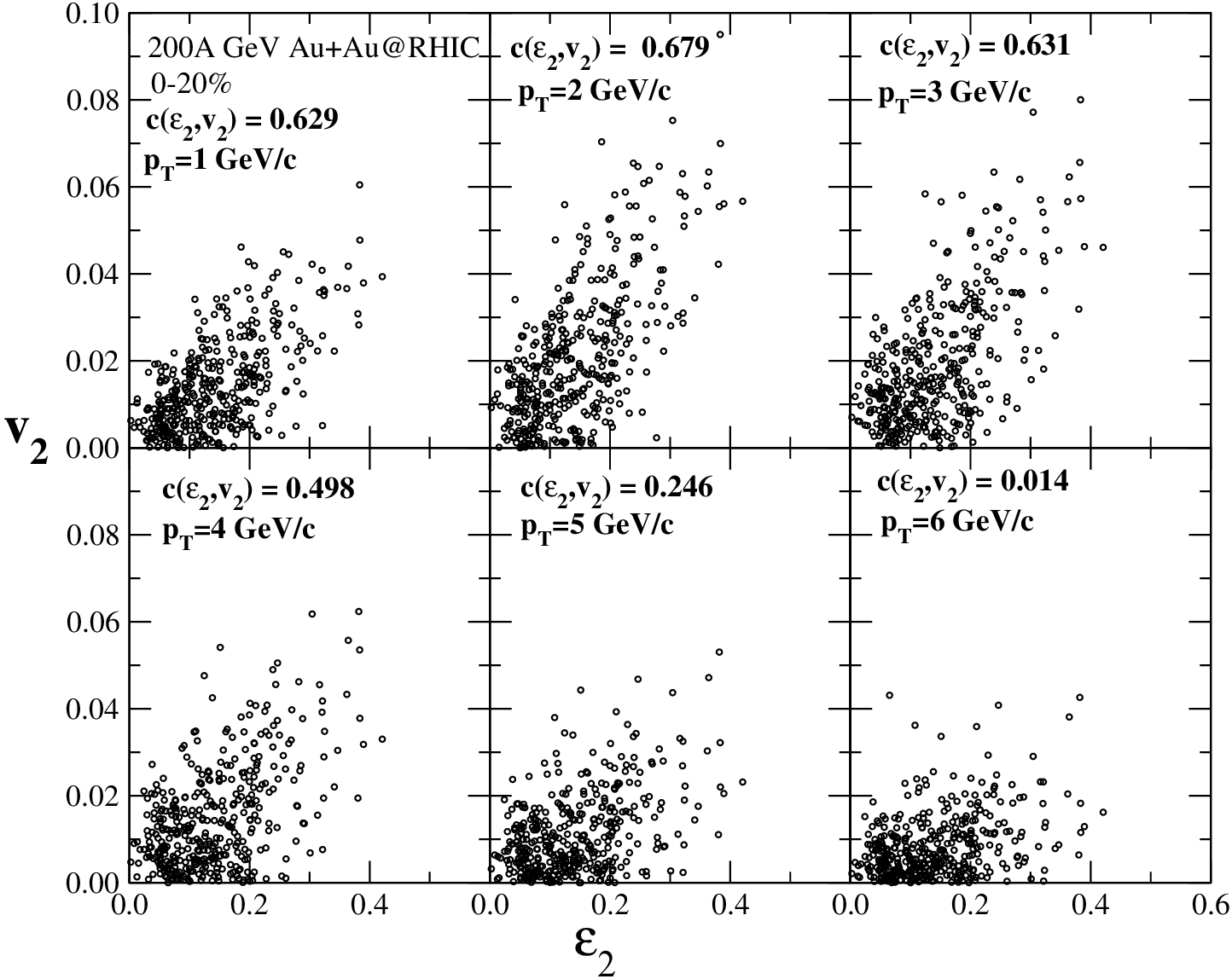}}
\centerline{\includegraphics*[width=7.0 cm]{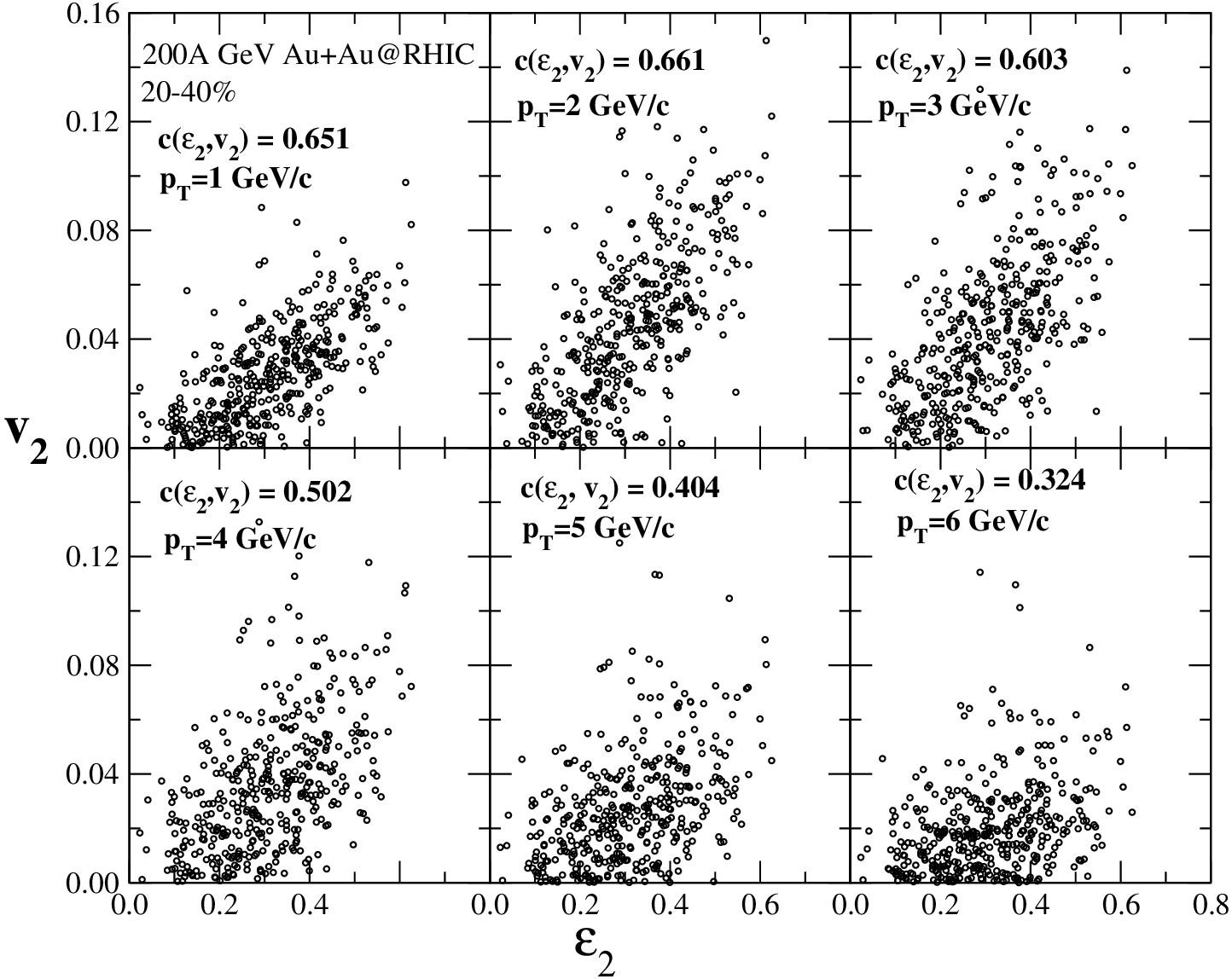}}
\centerline{\includegraphics*[width=7.0 cm]{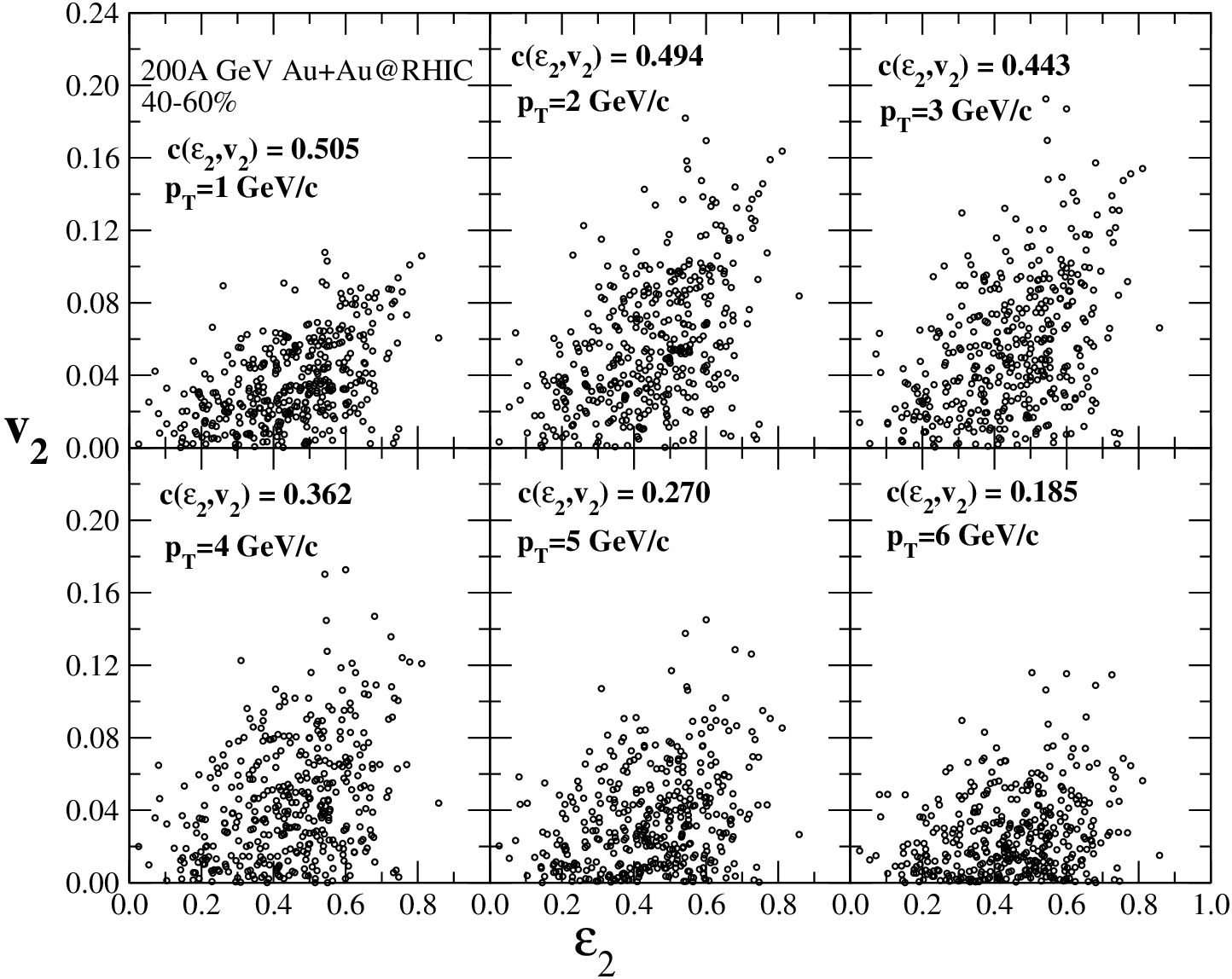}}
\caption{(Color online) Event-by-event distribution of thermal photon elliptic flow and initial spatial eccentricity at different $p_T$ values for 0--20\% (upper panel), 20--40\% (middle panel), and 40--60\% (lower panel) centrality bins for 200A GeV Au+Au collisions at RHIC.}
\label{fig6}
\end{figure}

\begin{figure}
\centerline{\includegraphics*[width=7.0 cm]{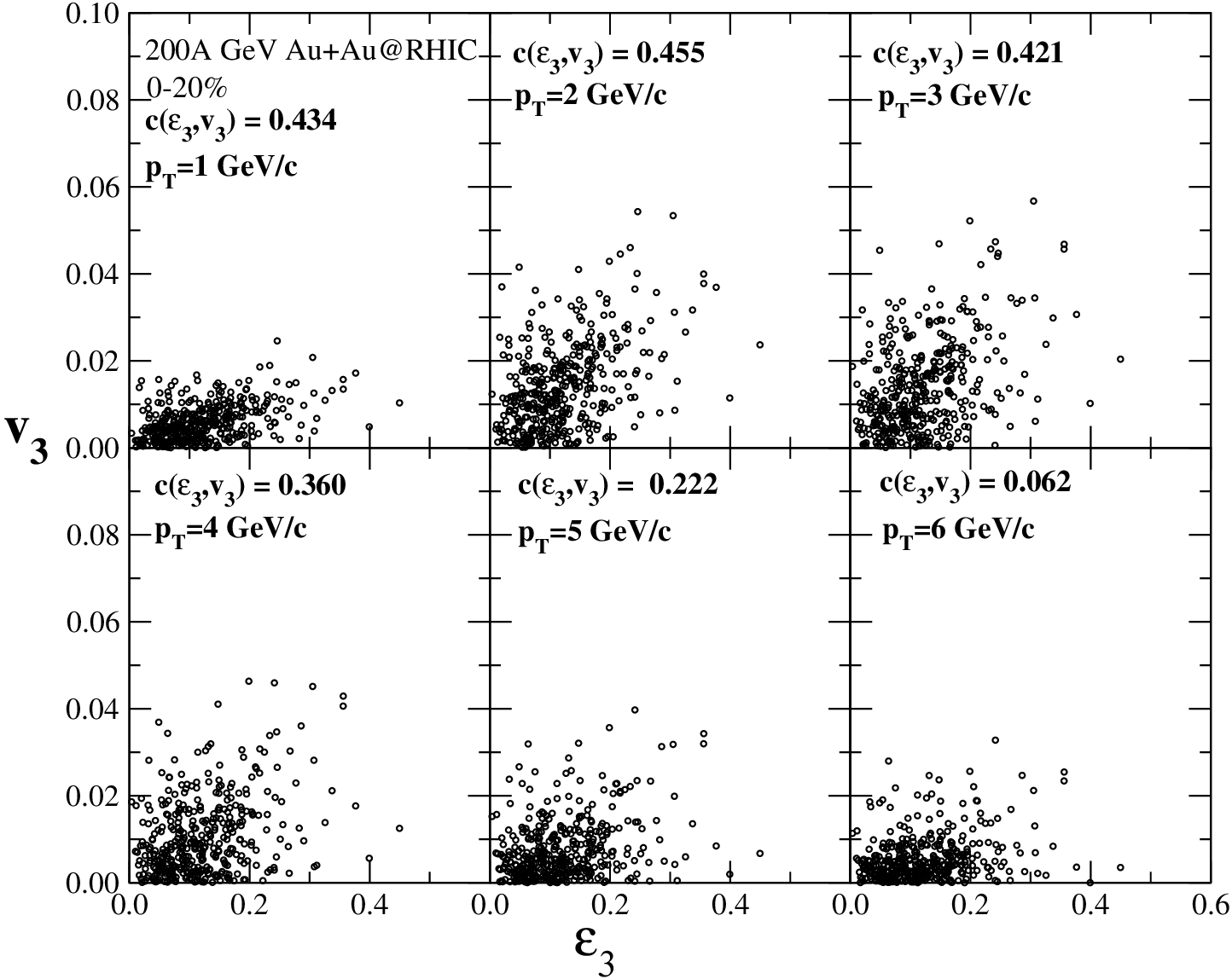}}
\centerline{\includegraphics*[width=7.0 cm]{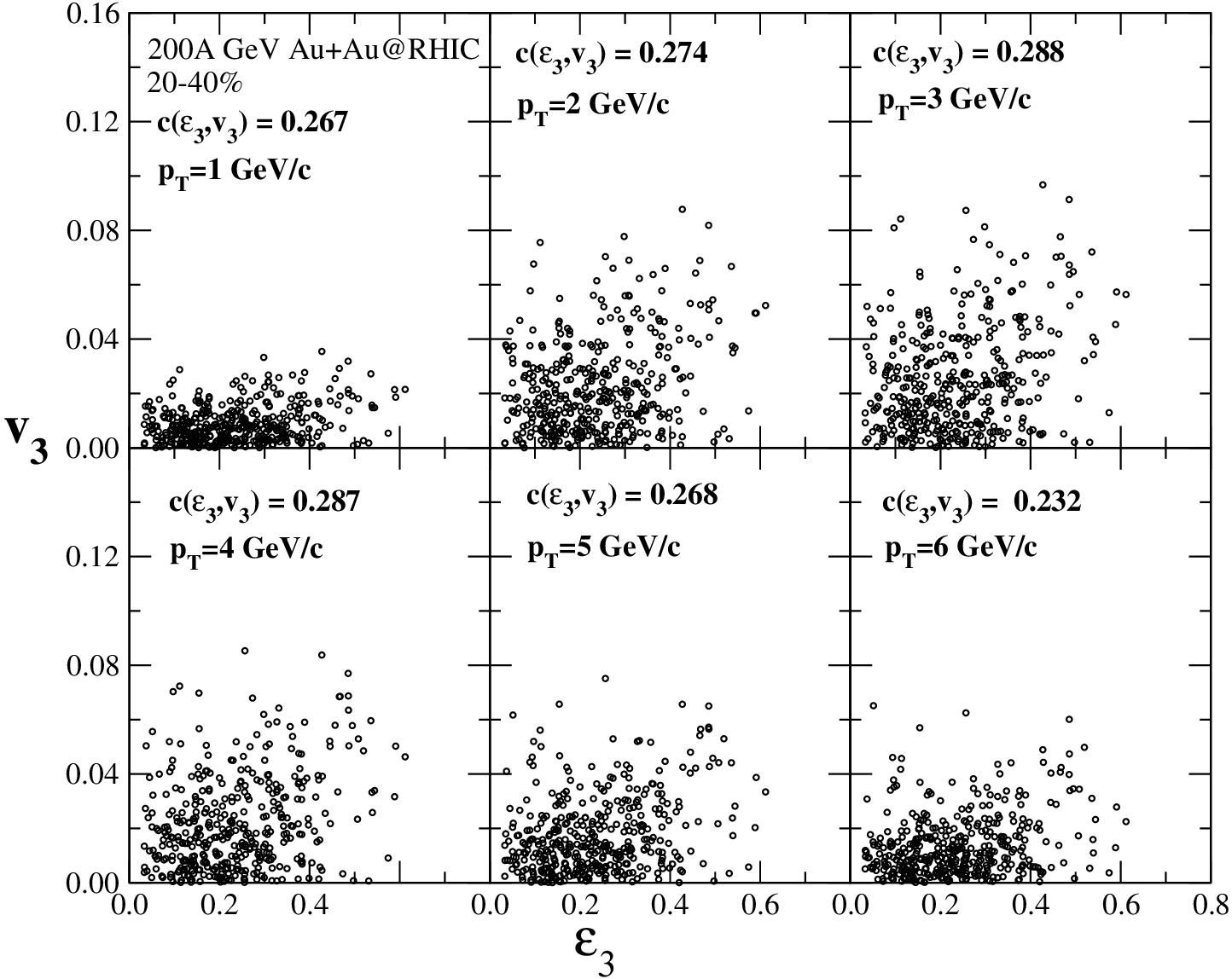}}
\centerline{\includegraphics*[width=7.0 cm]{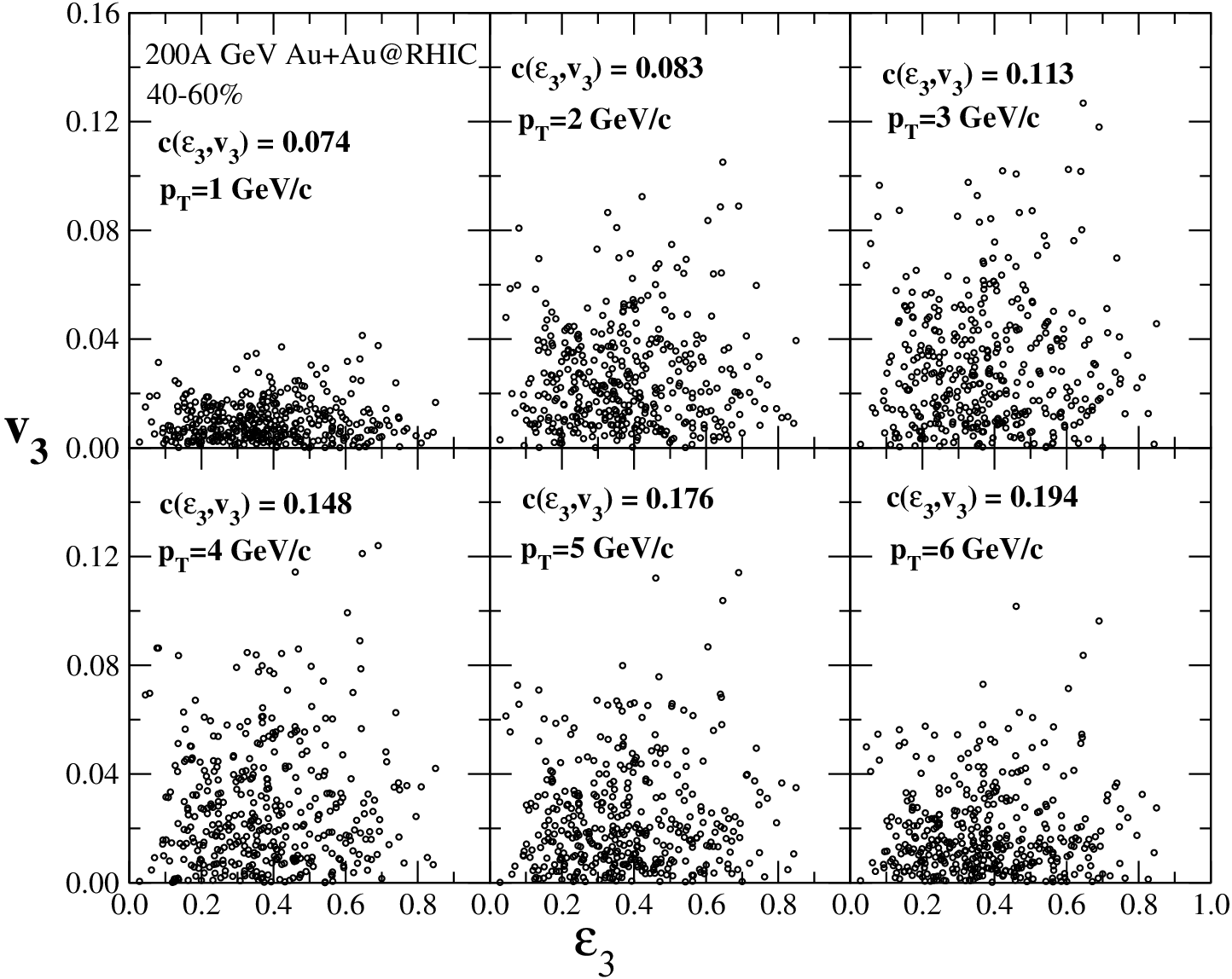}}
\caption{(Color online) Event-by-event distribution of thermal photon triangular flow and initial triangular eccentricity at different $p_T$ values for 0--20\% (upper panel), 20--40\% (middle panel), and 40--60\% (lower panel) centrality bins for 200A GeV Au+Au collisions at RHIC. }
\label{fig7}
\end{figure}

\begin{figure}
\centerline{\includegraphics*[width=7.0 cm]{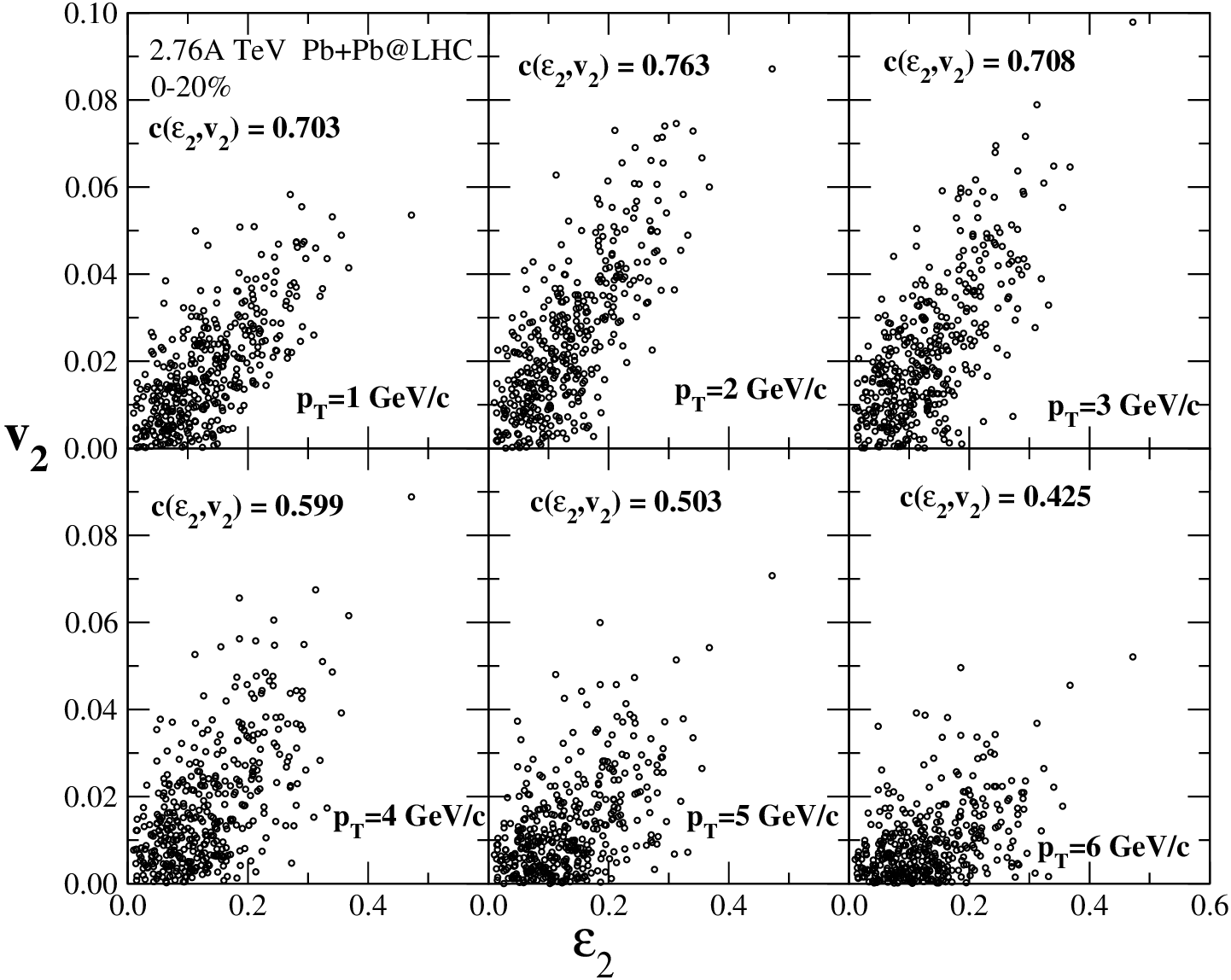}}
\centerline{\includegraphics*[width=7.0 cm]{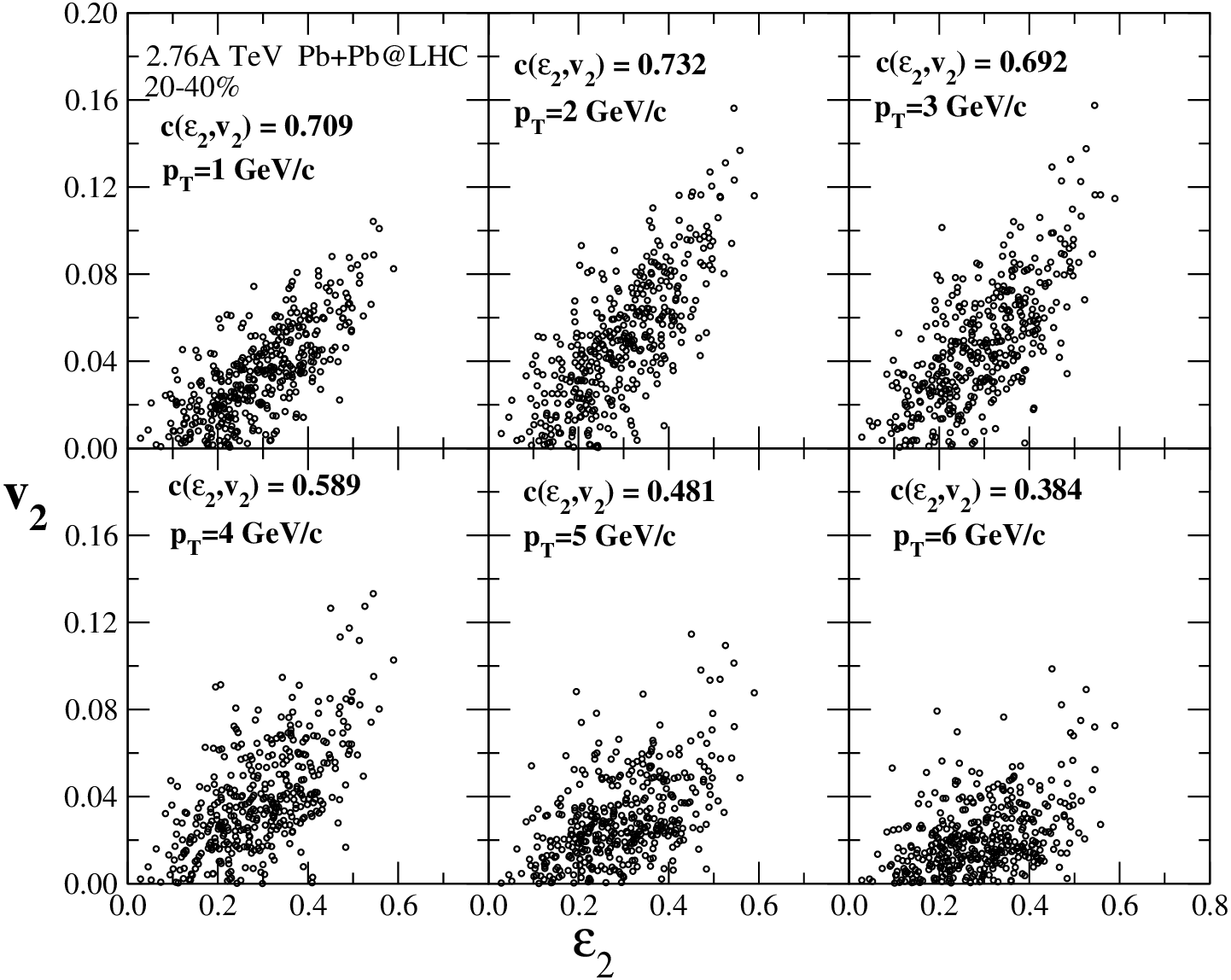}}
\centerline{\includegraphics*[width=7.0 cm]{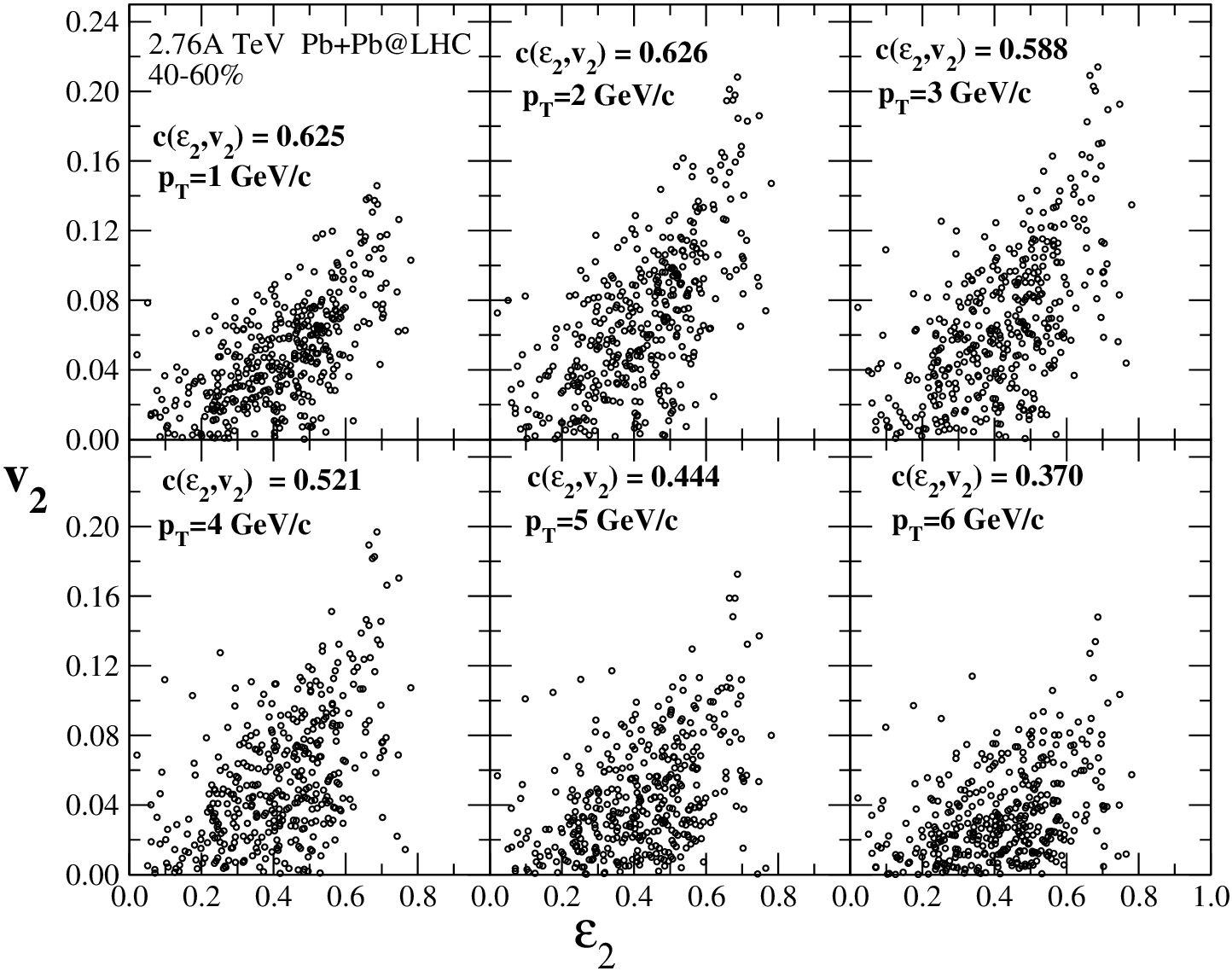}}
\caption{(Color online) Event-by-event distribution of thermal photon elliptic flow and initial spatial eccentricity at different $p_T$ values for 0--20\% (upper panel), 20--40\% (middle panel), and 40--60\% (lower panel) centrality bins for 2.76A TeV Pb+Pb collisions at LHC.}
\label{fig8}
\end{figure}

\begin{figure}
\centerline{\includegraphics*[width=7.0 cm]{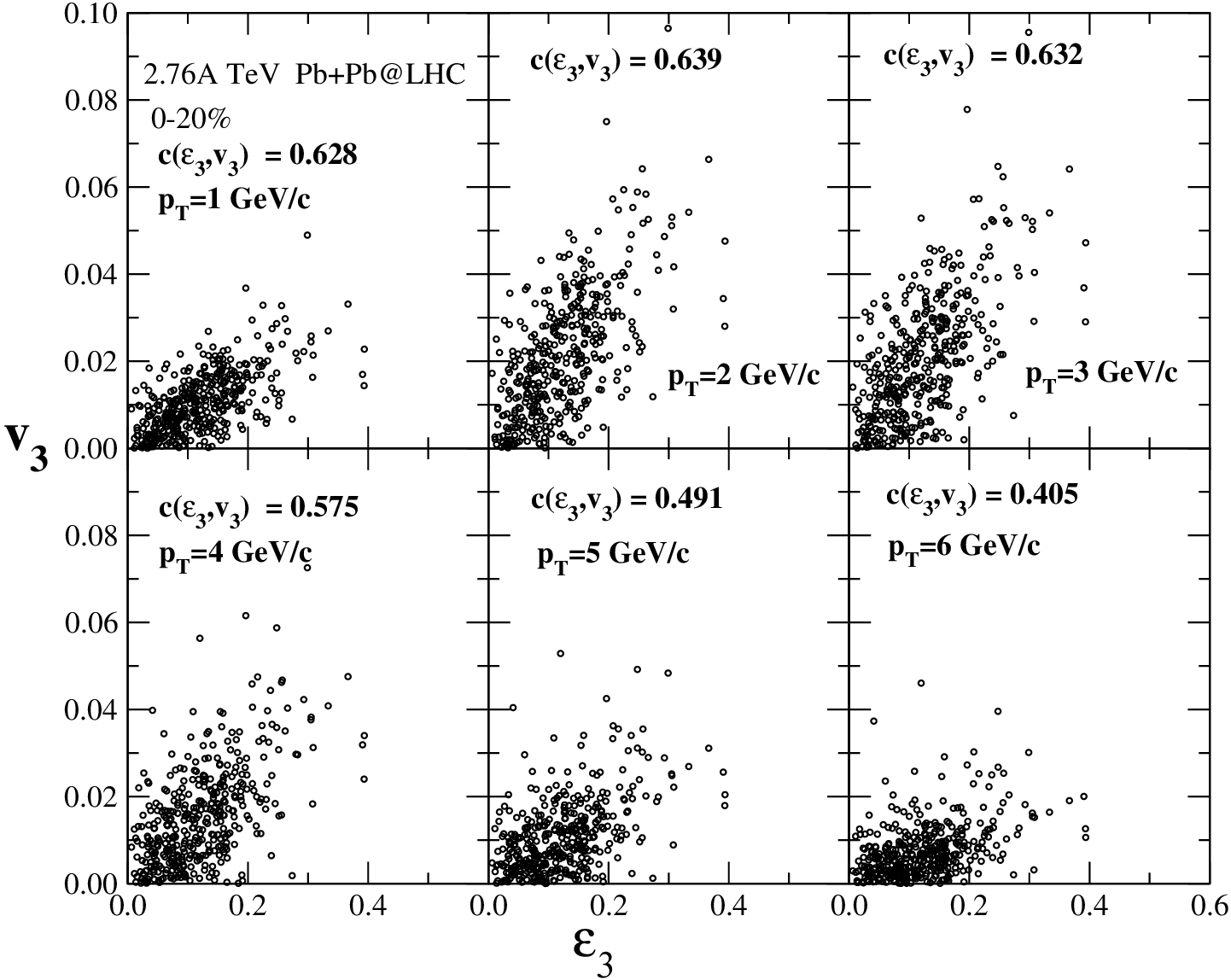}}
\centerline{\includegraphics*[width=7.0 cm]{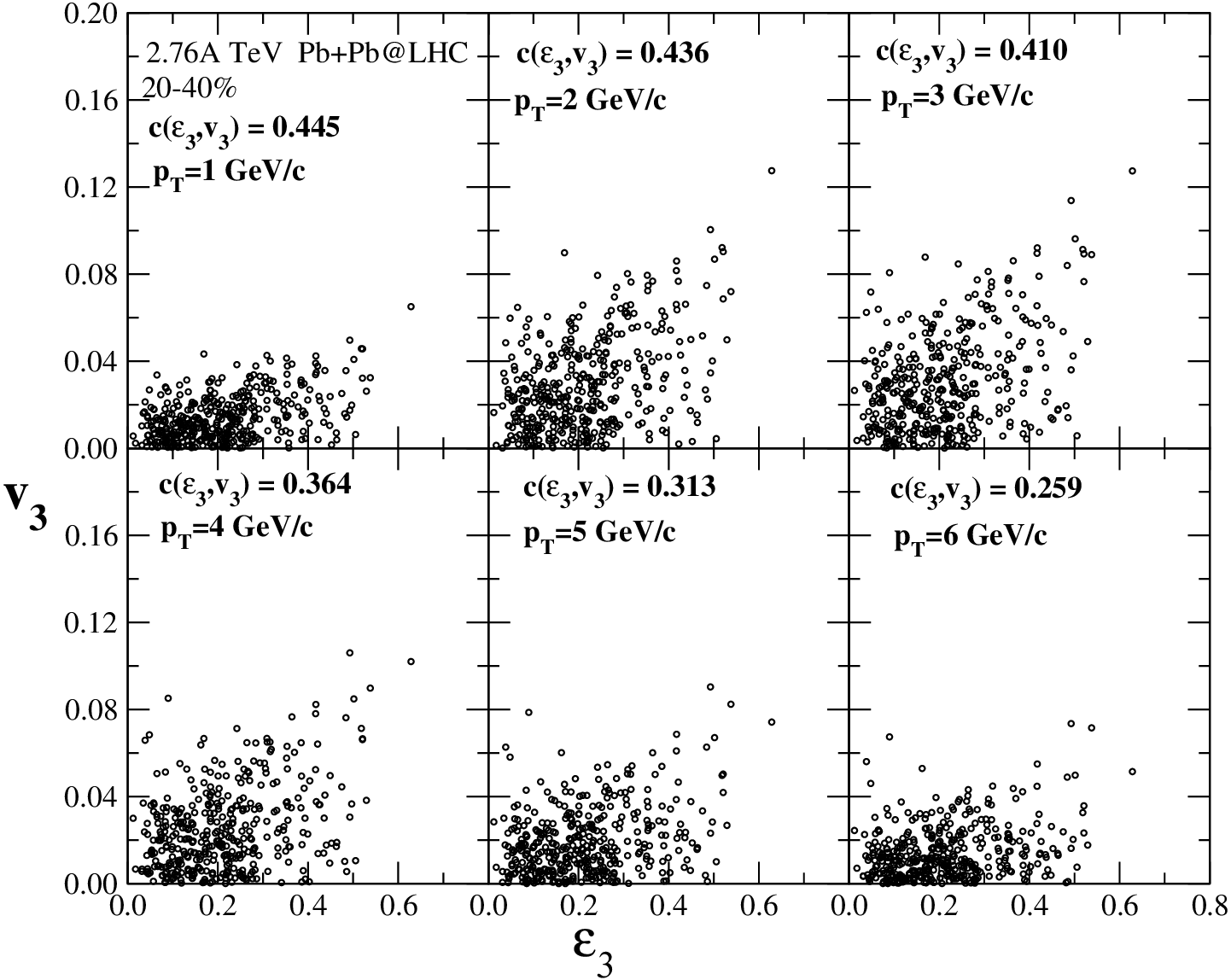}}
\centerline{\includegraphics*[width=7.0 cm]{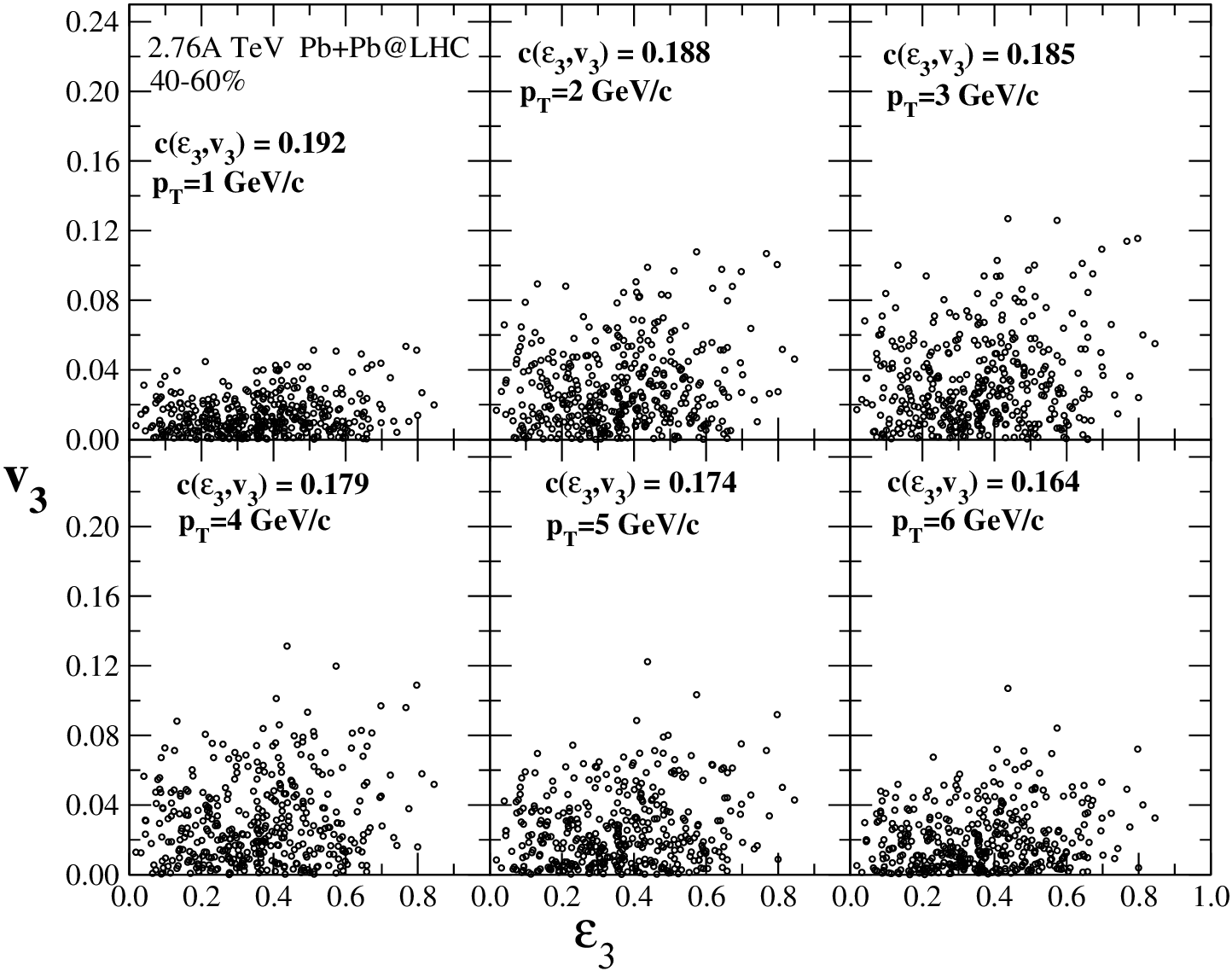}}
\caption{(Color online) Event-by-event distribution of thermal photon triangular flow and initial triangular eccentricity at different $p_T$ values for 0--20\% (upper panel), 20--40\% (middle panel), and 40--60\% (lower panel) centrality bins for 2.76A TeV Pb+Pb collisions at LHC.}
\label{fig9}
\end{figure}

\subsection{Correlation between $v_n$ and $\epsilon_n$}
It is quite well known and has been shown in earlier studies that the $p_T$ integrated elliptic flow of hadrons is proportional to the initial spatial eccentricity $\epsilon_2$ of the overlapping zone produced in collisions of heavy nuclei~\cite{uli}. The  final momentum anisotropy is considered to be the response of the  initial spatial anisotropy of  and the relation between $v_2$ and $\epsilon_2$ depends on the medium properties of the produced system. A recent interesting study shows that the event-by-event distributions of relative $v_n$ fluctuations are almost equal to the corresponding $\epsilon_n$ fluctuations and thus,  experimental determination of the relative anisotropy fluctuations of the initial state is possible~\cite{hannu_corr}. Same study shows that the linear correlation between elliptic low and corresponding  initial spatial anisotropy is close to 1, however,  the higher flow harmonics ($v_3$ and $v_4$) do not show significant linear correlation with the corresponding initial spatial anisotropies ($\epsilon_3$ and $\epsilon_4$). 

Photons are emitted from different stages of the system evolution and the $v_2$ is decided by the competing contributions from the quark matter and the hadronic matter phases. We also know that a high $p_T$ photon in the QGP phase originates mostly from the centre of the fireball, whereas a same  hight $p_T$ photon in the hadronic phase produced from the boundary. Thus, the correlation between the anisotropic flow parameters and the corresponding spatial eccentricities for photons is expected to be different than for hadron.

Next we study the correlation between the initial anisotropy and the corresponding  anisotropic flow parameters from their event by event distributions at different $p_T$ values. 

The correlation study using  $p_T$ integrated anisotropic flow and corresponding initial eccentricity would be valuable, however, this would be numerically expensive. We show that the results at different $p_T$ are equally important as the results can be understood in terms of the dominance of individual phases to the total photon $v_2$.

The linear correlation co-efficient $c(a,b)$ between quantities $a$ and $b$ is defined as
\begin{equation}
c(a,b)= \Big \langle  \frac{(a-\langle a\rangle_{\rm {evt}})(b-\langle b\rangle_{\rm {evt}})}{\sigma_a \sigma_b} \Big \rangle_{\rm {evt}} \ .
\end{equation}
where the averages $( \langle .. \rangle_{\rm evt})$ are taken over sufficiently large number of random events. $\sigma_a$ and $\sigma_b$ are the standard deviations of $a$ and $b$ respectively in the equation above. The linear correlation co-efficient can have value between +1 (linearly correlated) and -1 (anti-linearly correlated). The value $c(a,b)$ is zero when there is no linear correlation present between $a$ and $b$.

The event by-event distributions of $v_2$ and $\epsilon_2$ at RHIC are shown in Fig~\ref{fig6} and the distribution of $v_3$ and $\epsilon_3$ are shown in Fig.~\ref{fig7}. 
The correlation co-efficients $c (\epsilon_n, v_n)$ are calculated for different $p_T$ values for the three centrality bins.  The $\langle \epsilon_n \rangle$ and $\langle v_n \rangle$ are calculated by taking weight factor of impact parameter and particle yield respectively of the corresponding events. It is to be noted that a calculation of correlation co-efficients using larger number of events would give more reliable result. However, the results presented here are useful to know about the response of the initial geometry to the photon anisotropic flow.

The value of the co-efficient $c (\epsilon_2, v_2)$ is found to be larger ($\sim$ 0.5 to 0.7)  for $p_T$ values 1, 2, and 3 GeV/$c$. For higher $p_T$ values the liner correlation between $v_2$ and $\epsilon_2$ is found to be poor. The correlation is strongest for 20--40\% centrality bin at RHIC. 
The linear correlation between $\epsilon_3$ and $v_3$ is found to be much weaker than between $\epsilon_2$ and $v_2$. This confirms our understanding that the fluctuation driven larger transverse flow velocity plays a significant role apart from the initial geometry in determining the photon $v_3$. The correlation co-efficients $c (\epsilon_3, v_3)$ is largest for 0--20\% centrality bin and it decreases towards peripheral collisions.
We checked that the correlation co-efficient $c (\epsilon_n, v_n)$ calculated from event by-event distribution of photons from hadronic phase only is larger (not shown here) than the results presented in Fig.~\ref{fig6} and Fig.~\ref{fig7}. 

The distributions of photon anisotropic flow parameters and the initial eccentricities from Pb+Pb collisions at LHC are shown in Fig.~\ref{fig8} and Fig.~\ref{fig9}.
One can see that the correlation between $v_n$ and corresponding $\epsilon_n$ is stronger at LHC than the RHIC energy. The $c (\epsilon_2, v_2)$ varies in the range $\sim$ 0.6--0.8 for $p_T$ values 1, 2, and 3 GeV/$c$ for all three centrality bins at LHC. In addition, we that $c (\epsilon_3, v_3)$ values for 0--20\% centrality bin at LHC are significantly larger compared to RHIC.

\section{Summary and conclusions}
We calculate the elliptic flow and the triangular flow parameters of thermal photons as a function of $p_T$ from 200A GeV Au+Au collisions at RHIC for centrality bins 0--20\%, 20--40\% and 40--60\%  using an e-by-e hydrodynamic model with fluctuating initial conditions and compare the results with the anisotropic flow data from PHENIX Collaboration. 
The photon anisotropic flow coefficients are calculated with respect to the participant plane angle and for each centrality bin the results are obtained by taking average over flow parameters from 400 random events.

Our results from hydrodynamical model calculation are found to under-predict the experimental  direct photon $v_2$ and $v_3$ data by a large margin for all three centrality bins. We see by comparing the elliptic flow parameter at the three centrality bins that the $v_2(p_T)$ depends strongly on the collision centrality. Moving from central to peripheral collisions, the relative contribution from the hadronic phase to total photon flow increases. The photon $v_2$ from hadronic phase (only) is larger than the $v_2$ from QGP phase (only) and as a result total photon $v_2$ increases towards peripheral collisions. However, one can expect that photon $v_2$ might decrease with collisions centrality  $>$ 60\% as the overlapping zone itself become too small to generate large anisotropic flow. We can not check the results for ultra peripheral collisions as the present hydrodynamic framework is  believed to provide a reliable results upto a centrality bin 60\%. 
On the other hand, the triangular flow parameter do not show significant dependence on the collision centrality.
The photon $v_3$ is found to be maximum for the 20--40\% central collisions and the $v_3$ results for 0--20\% and 40--60\% centrality bins are found to be close to each other

In addition, we calculate the elliptic and triangular flow parameters of thermal photons at 2.76A TeV Pb+Pb collisions at the LHC for centrality bins 0--20\%, 20--40\%, and 40--60\%. The elliptic flow is found to increase slightly for all the centrality bins at LHC compared to RHIC. The centrality dependence of $v_2$ and $v_3$ at LHC is found to be similar to RHIC.

We calculate the $p_T$ spectrum and anisotropic flow parameters of thermal photons from a relatively smaller system formed in collisions of Cu nuclei at 200A GeV at RHIC and for 0--20\% centrality bin. The photon production is found to be significantly smaller for Cu+Cu collisions than for Au+Au collisions at RHIC. However, the effect of fluctuations is found to be more pronounced in the anisotropic flow results for the smaller system. We see a larger elliptic and triangular flow parameter of thermal photons from Cu+Cu collisions compared to Au+Au collisions.

Next we study the correlation between the initial spatial anisotropies and the final momentum space anisotropies at different $p_T$ values of thermal photons from their e-by-e distribution. 
The correlation between $\epsilon_n$ and $v_n$ is found to be weaker for larger values of $n$. The correlation co-efficient is found to be larger at LHC than at RHIC.

\begin{acknowledgments} 
We acknowledge the computer facility of Drona and  Prafulla clusters of VECC. DKS gratefully acknowledges the grant of Raja Ramanna Fellowship by the Department of Atomic Energy, India.
\end{acknowledgments}

\end{document}